\documentclass[
    aps,
    prl,
    reprint,
    superscriptaddress,
    nolongbibliography,
    nobibnotes,
]{revtex4-2}

\usepackage[dvipsnames, usenames]{xcolor}
\definecolor{linkcolor}{rgb}{0.1, 0.5, 0.7}

\usepackage[
colorlinks=true,
linkcolor=linkcolor,
citecolor=linkcolor,
filecolor=linkcolor,
urlcolor=linkcolor,
pdfusetitle,
]{hyperref}

\usepackage{amssymb}
\usepackage{amsmath}
\usepackage{physics}
\usepackage{orcidlink}
\usepackage{acronym}
\usepackage{dsfont}
\usepackage[OT1]{fontenc}
\usepackage{xspace}
\usepackage{booktabs}
\usepackage{xparse}
\usepackage{bm}
\usepackage{eso-pic}

\usepackage{diagbox}

\newcommand{\red}[1]{{#1}}

\NewDocumentCommand{\ct}{e{_}e{^}}{%
  \ensuremath{%
    \cos
    \IfValueT{#2}{^{#2}}%
    \theta
    \IfValueT{#1}{_{#1}}%
  }\xspace%
}

\newcommand{\chie}{\ensuremath{\chi_\mathrm{eff}}\xspace}
\newcommand{\chip}{\ensuremath{\chi_\mathrm{p}}\xspace}
\newcommand{\chib}{\ensuremath{\chi_\mathrm{eff, p}}\xspace}
\newcommand{\s}{\ensuremath{s}\xspace}

\newcommand{\pf}{\ensuremath{g}\xspace}
\newcommand{\pfp}{\ensuremath{\gamma}\xspace}

\newcommand{\moddef}{\textsc{Default}\xspace}
\newcommand{\modbiv}{\textsc{Bivariates}\xspace}

\newcommand{\Var}[1]{\mathrm{Var}\,#1}
\newcommand{\mean}[1]{\left\langle #1 \right\rangle  }

\newcommand{\lnb}{\ensuremath{\Delta \ln Z}\xspace}

\newcommand{\desq}[1]{\delta^2\!\left(#1\right)}

\newacro{MC}{Monte Carlo}
\newacro{IID}{Independent and Identically Distributed}
\newacro{NID}{Non-independent but Identically Distributed}
\newacro{BBH}{binary black hole}
\newacro{BH}{black hole}
\newacro{GW}{gravitational-wave}
\newacro{LVK}{LIGO--Virgo--KAGRA}
\newacro{PPD}{posterior population distribution}
\newacro{KL}{Kullback--Leibler}
\newacro{KDE}{Kernel Density Estimate}

\newcommand{\ligo}{\affiliation{LIGO Laboratory, Massachusetts Institute of Technology, Cambridge, MA 02139, USA}}
\newcommand{\mki}{\affiliation{Kavli Institute for Astrophysics and Space Research, Massachusetts Institute of Technology, Cambridge, MA 02139, USA}}
\renewcommand{\mit}{\affiliation{Department of Physics, Massachusetts Institute of Technology, Cambridge, MA 02139, USA}}

\begin{document}

\AddToShipoutPictureFG*{%
  \AtPageUpperLeft{%
    \put(5,-50){\includegraphics[width=4cm]{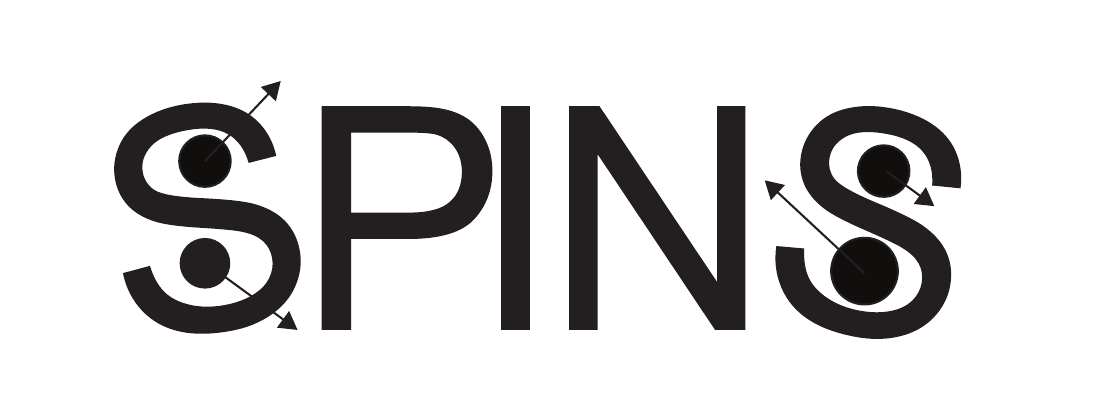}}%
  }%
}

\title{Binary-black hole spin population results may be driven by prior degeneracies}

\author{Noah E. Wolfe\,\orcidlink{0000-0003-2540-3845}}
\email{newolfe@mit.edu}
\ligo\mki\mit

\author{Asad Hussain\,\orcidlink{0000-0003-3491-5439}}
\affiliation{Center for Computational Astrophysics, Flatiron Institute, NY}

\author{Jack Heinzel\,\orcidlink{0000-0002-5794-821X}}
\ligo\mki\mit

\author{Salvatore Vitale\,\orcidlink{0000-0002-0147-0835}}
\ligo\mki\mit

\collaboration{Society of Physicists Interested in Non-aligned Spins, SPINS}
\homepage{https://sites.mit.edu/spins}

\date{\today}

\begin{abstract}
Gravitational waves carry information on the spins of merging binary-black holes.
The orientations of their spins relative to their orbits (``tilts'')---while challenging to measure---differentiate the astrophysical formation mechanisms
by which merging black hole binaries may form.
Multiple analyses have reported tentative evidence that black hole spins
are preferentially oriented in the plane of the binary orbit.
We show that these results are likely extrapolated from more easily measurable ``effective spin'' parameters
which characterize the gravitational-wave inspiral of black hole binaries
but may not uniquely constrain their astrophysical formation.
In particular,
we reproduce a peak in the spin tilt distribution
with information about the marginal effective spin distributions alone.
We propose a geometric picture
to compare constraints on effective spins
to standard population modeling assumptions,
suggesting that current inferences of a preferred spin orientation are spurious.

\end{abstract}

\maketitle

\textit{Introduction}---The fifth \ac{LVK} \cite{LIGOScientific:2014pky, VIRGO:2014yos, KAGRA:2020tym} catalog of \ac{GW} sources
includes 259 \ac{BBH} mergers \cite{LIGOScientific:2026sit, LIGOScientific:2026wfs}.
By measuring the population distribution of their source properties,
we can constrain the astrophysical channels by which these systems formed.
\Ac{BH} spins in particular have long been considered a key differentiator between different \ac{BBH} formation histories.
For example, hierarchical generations of mergers may tend to produce
rapidly spinning \acp{BH} \cite{Berti:2008af, Gerosa:2021mno}.
Binaries evolving in dense stellar environments
may tend to have spins randomly misaligned relative to their orbit \citep{2016ApJ...832L...2R, 2022ApJS..258...22R}.
In comparison, binary formation in the field may tend to yield \acp{BH}
spinning slowly and aligned with the binary orbit
\cite{Kalogera:1999tq, Gerosa:2018wbw, Belczynski:2017gds, Fuller:2019sxi}.

Recent studies have found that most merging \acp{BH} tend to be slowly spinning,
although there may be subpopulations of binaries where at least
one \ac{BH} is rapidly spinning \cite{Godfrey:2023oxb, Li:2023yyt, Hussain:2024qzl, Banagiri:2025dmy, Farah:2026jlc, Wolfe:2026meb, Galaudage:2026opk, Guttman:2026cnv}.
Multiple analyses have found tentative evidence
for a preferred \ac{BH} spin orientation of $\sim75^\circ\,$--$\,90^\circ$ relative to the orbital angular momentum \cite{Godfrey:2023oxb, Golomb:2022bon, Callister:2023tgi, Vitale:2022dpa, Stegmann:2025zkb, Li:2025iux, LIGOScientific:2025pvj, LIGOScientific:2026ctl, Guttman:2025jkv}.
While not yet required by the data---and model dependent \cite{Vitale:2022dpa, Wolfe:2026meb}---if a preference for in-plane spins is confirmed,
it would challenge canonical theories of \ac{BBH} formation
and suggest other channels
such as binary evolution in the presence of a tertiary companion \cite{Antonini:2017tgo, Rodriguez:2018jqu, Liu:2018nrf, Yu:2020iqj, Su:2020vda}.
See Ref.~\cite{Biscoveanu:2026ikx}
for a review of theoretical predictions
and current \ac{GW} spin measurements.

Simulation studies also consistently reproduce tentative evidence of a preference for in-plane spins---even when the astrophysical population
favors \ac{BBH} spins aligned with their orbits \cite{Vitale:2025lms}
or has no preferred spin orientation \cite{Wolfe:2026meb}.
In principle,
the spin inferences obtained with real and simulated data
could represent Poisson fluctuations in catalog membership \cite{Vitale:2025lms, Wolfe:2026meb, Corelli:2026thw}.
However, support for in-plane \ac{BH} spins in real data has persisted
as the \ac{GW} catalog has grown from 69 \cite{Vitale:2022dpa, KAGRA:2021duu} to 259 \cite{LIGOScientific:2026ctl} events.
These results are somewhat surprising given the difficulty of measuring \ac{BBH} spin orientations \cite{vanderSluys:2007st, vanderSluys:2009bf, Pankow:2016udj, Vitale:2016avz}.

We characterize \acp{BBH} 
by their masses $m_{1,2}$ with ratio $q = m_2 / m_1 \leq 1$,
the magnitudes $a_{1,2}$ of their spins,
and the cosine of the angle (the ``tilt'') between their spins and the orbital angular momentum $\ct_{1,2}$.
The primary (secondary) \ac{BH} is the more (less) massive of the pair.
Astrophysical \acp{BBH} are also characterized by their redshift, though not a focus of this work.
Typically, we measure best the effective spins of \acp{BBH} \cite{Purrer:2015nkh, Vitale:2016avz, Shaik:2019dym, Pratten:2020igi, Green:2020ptm, Biscoveanu:2021nvg, Krishnendu:2021cyi, Miller:2025eak}
which are leading-order parameters of \ac{GW}-driven inspiral \footnote{
See Refs.~\cite{Gerosa:2020aiw, Thomas:2020uqj} for extensions of \chip which more accurately capture \ac{GW} inspiral spin precession morphology.
}:
the aligned effective spin \cite{Damour:2001tu, Racine:2008qv, 2014LRR....17....2B},
\begin{equation} \label{eq:chieff-def}
    \chie = \frac{a_1 \ct_1 + q \, a_2 \ct_2}{1 + q} \, ,
\end{equation}
and the precessing effective spin \chip \cite{Schmidt:2014iyl},
\begin{equation} \label{eq:chip-def}
    \chip = \max{\left( a_1 \sin \theta_1, q\frac{4 q + 3}{3q + 4} a_2 \sin \theta_2 \right)} \, .
\end{equation}
We collectively refer to the component spin magnitudes and tilts as $\s = (a_1, a_2, \ct_1, \ct_2)$.

There is information beyond the effective spins contained in some individual events \cite{plunkett2026prep}.
However,
it is not yet clear if this information combines
at the population level
to provide constraints on the component spin distribution \cite{Miller:2024sui}.
Rather,
population-level conclusions about component spins could reflect
a model extrapolation
of constraints on the effective spin distribution.
The effective spins are lossy representations of the four component spins.
Thus,
conclusions about the \ac{BH} spin distribution
that are derived from a combination of effective spins
and model priors
may not carry
a unique astrophysical interpretation.

In this \textit{Letter},
we develop a method for identifying model extrapolation in spin population inference
which we apply to the fifth Gravitational-Wave Transient Catalog (GWTC-5).
We find that knowledge of the effective spin distribution alone
is sufficient to reproduce a preferred \ac{BH} tilt
when viewed through standard spin population model assumptions.
We recommend against the use of models which assume that primary and secondary spins are identically distributed.
Our results even suggest that models which neglect correlations between spin magnitudes and tilts are overly strong relative to the information contained in current \ac{GW} catalogs.
Thus, caution is required when interpreting astrophysical claims made based on \ac{BBH} component spins.

\textit{Hierarchical Inference}---For \acp{BBH} with source parameters $\vartheta$
(including spins, masses, and redshift)
we adopt a functional form---the population model---$p(\vartheta \mid \lambda)$ with parameters $\lambda$
to model the population distribution.
Then, the likelihood of a GW event with data $d$ is
\begin{align} 
    p(d \mid \lambda) &= \int \dd \vartheta \, p(d \mid \vartheta) \, p(\vartheta \mid \lambda) \\
    &= \int \dd q \, \dd s \, p(d \mid s, q) \, p(s \mid \lambda) \, p(q \mid \lambda) \, , \label{eq:comp-spin-like}
\end{align}
where from the first to second line we carry out the integral over
source properties besides $s$ and $q$, e.g., primary mass and redshift.
Note that $p(d \mid \vartheta)$ encodes our model of \ac{GW} physics (e.g., gravitational waveform approximant and detector noise)
while $p(d \mid \lambda)$
weights data according to the astrophysical population
sourcing those data.
Hierarchical inference \cite{Loredo:2001rx, 2019MNRAS.486.1086M, Vitale:2020aaz} evaluates $p(d \mid \lambda)$ for each event in the catalog $\{ d \}$
and combines the results while accounting for selection effects to obtain the population likelihood $p(\{ d \} \mid \lambda)$.
Adopting a population prior $p(\lambda)$ yields the population posterior $p(\lambda \mid \{ d \}) \propto p(\{ d \} \mid \lambda) p(\lambda)$.

We analyze the \acp{BBH} in GWTC-5.
Event selection criteria, detector sensitivity estimation, and waveform choices
are described in the Supplemental Material.
We estimate the population likelihood
by \ac{MC} integration
and exclude $\lambda$ for which this 
estimator's variance is higher than acceptable \cite{Tiwari:2017ndi, Essick:2022ojx, Talbot:2023pex, Heinzel:2025ogf}.
See the End Matter for details.
We apply the same cuts when
showing population priors.
\red{
Note that these cuts asymmetrically
modify the component spin prior densities;
see the Supplemental Material.
}

\textit{Restricting to effective spins}---For a given choice of component spin population model $p(s \mid \lambda)$
we want to determine if constraints
on the component spin distribution
are directly informed by \ac{GW} catalogs
or instead through constraints on the effective spin distribution.
This goal is accomplished for per-event spin measurements through prior conditioning \cite{Gangardt:2022ltd, plunkett2026prep}.
We formalize this method for population-level inference by restricting the population model and likelihood to only include information about effective spins.

We modify the likelihood in Eq.~\eqref{eq:comp-spin-like}
by marginalizing away spin degrees of freedom beyond the effective spins \chib.
Then, the likelihood is
\begin{equation} \label{eq:eff-spin-like}
    \int \dd q \, \dd \chi_\mathrm{eff, p} \, p(d \mid \chi_\mathrm{eff, p}, q) \, p(\chi_\mathrm{eff, p} \mid q, \lambda) \, p(q \mid \lambda) \, ,
\end{equation}
where the $\chib$ distribution is conditioned on $q$ because
of how effective spins are defined in Eqs.~\eqref{eq:chieff-def} and \eqref{eq:chip-def}.
Remaining spin degrees of freedom beyond \chib are assumed to be uniformly distributed.
Typically when adopting the likelihood defined in Eq.~\eqref{eq:eff-spin-like} over Eq.~\eqref{eq:comp-spin-like},
analysts write the population model explicitly in terms of the effective spins.
This defines a model with different parameters, and over a different functional space, than component spin population models.

Instead, we map a component spin model to effective spins, which we call the \textit{pushforward} of the component spin model,
\begin{equation} \label{eq:pf}
    p(\chib \mid q, \lambda) = \int \dd s \, p(s \mid \lambda) \, p(\chib \mid s, q) \, ,
\end{equation}
where $p(\chib \mid s,q ) = \desq{ \chib(s, q) - \chib }$\red{; cf. App.~C.5 in Ref.~\cite{Essick:2025zed}.}
The pushforward cannot be evaluated analytically for all choices of $p(s \mid \lambda)$ \footnote{When $p(s \mid \lambda) \propto 1$ we recover the Jacobian derived by Refs.~\cite{Callister:2021gxf, Iwaya:2024zzq}. See Ref.~\cite{Farr:2017uvj} for other analytic pushforwards.}
so we estimate Eq.~\eqref{eq:pf} by \ac{MC} simulation \footnote{See similarly Ref.~\cite{Callister:2020vyz}
which inferred the \chie distribution
with a population model estimated for
 a physically motivated \ac{MC} simulation of component spins.}.
Given $\lambda$ we draw
samples $\{ s, q \} \sim p(s \mid \lambda)p(q \mid \lambda)$
transformed to samples $\{ \chib \}$ via Eqs.~\eqref{eq:chieff-def} and \eqref{eq:chip-def}.
We approximate the pushforward 
by fitting these samples
with a density \pf
with parameters \pfp.
At fixed $\lambda$, \pfp are chosen
such that $\pf(\chib \mid \pfp) \approx p(\chib \mid q, \lambda)$ for all $q$.
The pushforward approximant
fits \chie and \chip independently;
$\{ \chie \}$ with a generalized Gaussian (allowing heavier or lighter tails than a standard Gaussian)
and $\{ \chip \}$ with a Beta distribution.
See the End Matter for implementation details.

We denote the approximate pushforward density $\pf(\chib \mid \pfp(\lambda)) \approx p(\chib \mid q, \lambda)$ for all $q$.
In turn we approximate Eq.~\eqref{eq:eff-spin-like} as
\begin{equation} \label{eq:approx-eff-like}
\begin{aligned}
    p(d \mid \lambda, \mathrm{eff}) = \int \dd q \, \dd &\chib \, p(d \mid \chib, q) \, \times \\
    &\times \pf(\chib \mid \pfp(\lambda)) \, p(q \mid \lambda) \, ,
\end{aligned}
\end{equation}
where the label ``eff''
distinguishes this likelihood from Eq.~\eqref{eq:comp-spin-like}.
Evaluating Eq.~\eqref{eq:approx-eff-like} for all data in a catalog,
we obtain the hierarchical population likelihood $p(\{ d \} \mid \lambda, \mathrm{eff})$.
When analyzing \ac{GW} catalogs with the pushforward model
we adopt the same prior $p(\lambda)$ as
used for the corresponding component spin analysis.
Then, we obtain the 
\textit{restricted}
posterior $p(\lambda \mid \{ d \}, \mathrm{eff}) \propto p(\{ d \} \mid \lambda, \mathrm{eff}) p(\lambda)$
in contrast to the \textit{full} posterior $p(\lambda \mid \{ d \})$.

We compare the restricted and full posterior constraints
on the \ac{BBH} component spin distribution.
Where these constraints coincide,
we conclude that
the component spin population model
has extrapolated from the effective spin distribution
rather then solely reflecting information
contained in the \ac{GW} catalog.

We quantify information gain from the restricted posterior to the full posterior
with log Bayes factors \lnb;
evidences are estimated with nested sampling
and we report standard errors on \lnb.
We also propagate \ac{MC} uncertainty in the log-likelihood estimator
into \lnb; see the End Matter for details.
When that evidence is weak, cf. Ref.~\cite{kass1995bayes},
we conclude that the catalog
does not inform the spin distribution
beyond the effective spins.

\begin{figure}
    \centering
    \includegraphics[width=0.75\linewidth]{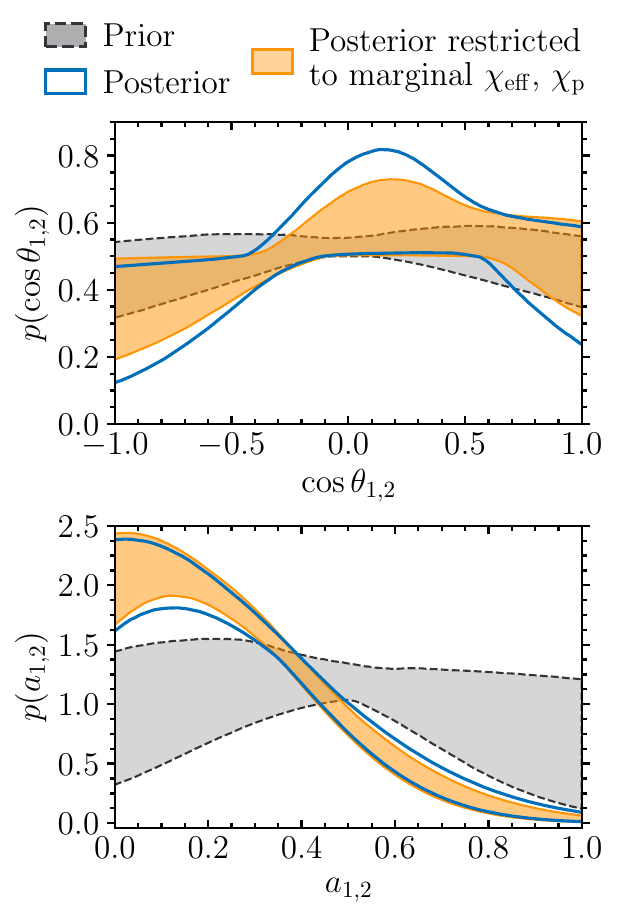}
    \caption{
    Prior (grey, dashed) and full posterior (blue) spin distributions under the \moddef model,
    compared to the posterior restricted to the marginal \chie, \chip distributions (yellow).
    Lines or fill enclose 90\% credible levels.
    }
    \label{fig:default-ppds}
\end{figure}

\begin{figure*}
    \centering
    \includegraphics[width=0.98\linewidth]{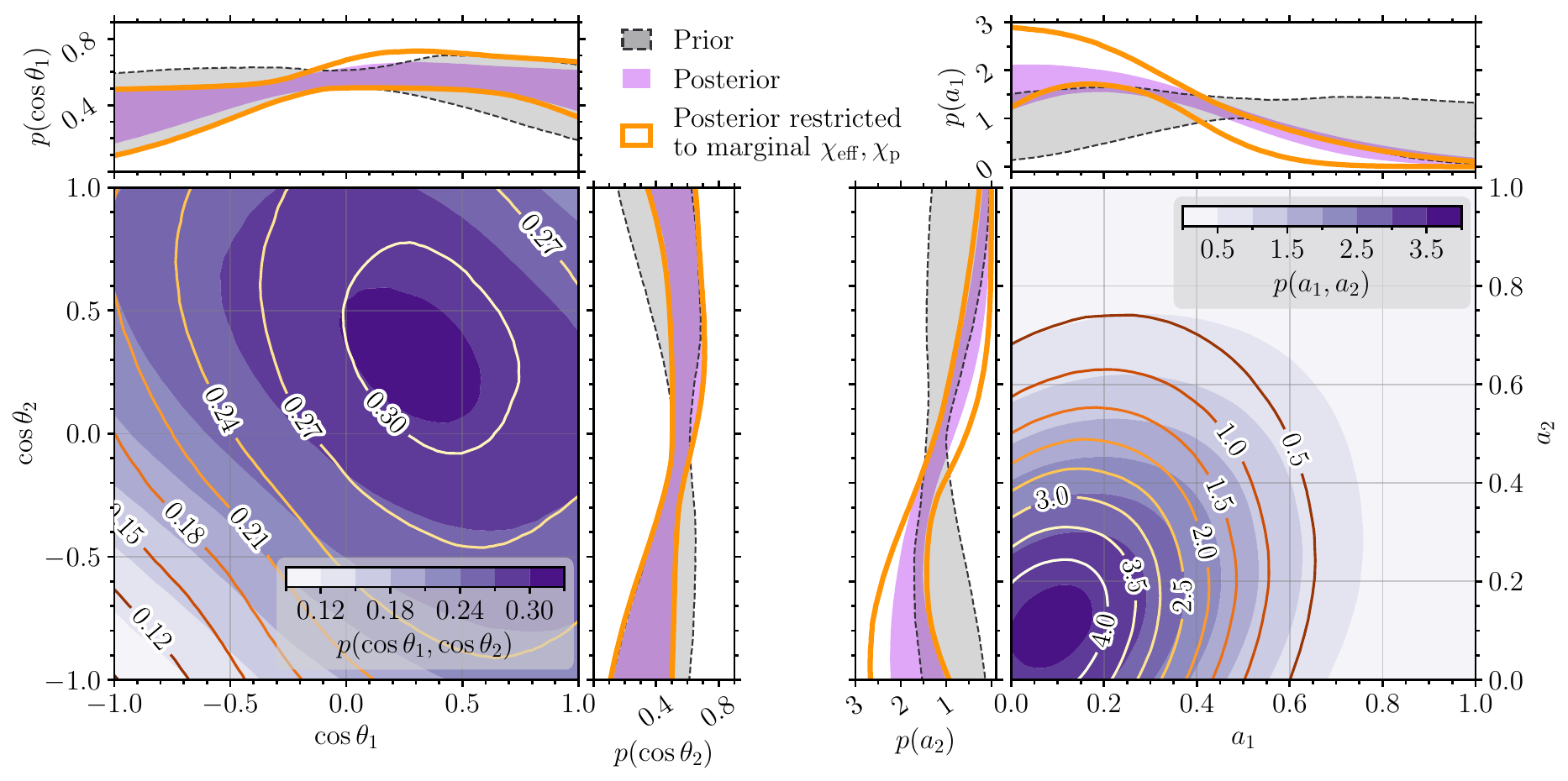}
    \caption{
    Restricted (yellow) and full posterior (purple) population distributions for spin tilts (left) and magnitudes (right)
    under the \modbiv model.
    Middle panels show the median population distribution
    while marginals show the 90\% credible region.
    Marginal priors are in grey. 
    }
    \label{fig:biv-ppds-joint}
\end{figure*}

\textit{Models \& Results}---We infer the \ac{BBH} spin distribution
with two different models
for the component spins.
The \moddef model adopts the preferred parametric model for component spins from Ref.~\cite{LIGOScientific:2026ctl}.
We model $a_{1,2}$ as \ac{IID}
according to a truncated Gaussian
and $\ct_{1,2}$ with a mixture between isotropic and truncated Gaussian distributions \cite{Vitale:2015tea, Talbot:2017yur, Vitale:2022dpa}.
Formally, $\ct_1$ and $\ct_2$ are \ac{NID}; either they are both drawn from the isotropic distribution or both from the Gaussian, but their marginals are identical.
In the \modbiv model, we relax the assumption that primary and secondary spins are identically distributed.
We model $a_{1,2}$ and $\ct_{1,2}$ each according to bivariate truncated Gaussians \cite{Hussain:2024qzl}.
Both analyses model masses and redshift according to the preferred parametric model from Ref.~\cite{LIGOScientific:2026ctl}.
We repeat each analysis
with the corresponding effective spin pushforward model.
\red{Population models, priors, and analysis settings are detailed in the Supplemental Material.}

Results obtained with the \moddef model are shown in Fig.~\ref{fig:default-ppds}.
In the top panel, we find that the analysis
restricted to the marginal \chib distribution reproduces support for a peak in the tilt distribution
at the same location as in the full component spin analysis,
at $\ct_{1,2} \sim 0$--0.2,
and with similar amplitude.
We quantify the peak location and amplitude with
the tilt population mean $\mean{\ct_{1,2}}$ and variance; note that an isotropic tilt distribution has $\Var{\ct_{1,2}}=1/3$.
The restricted posterior has
$\mean{\ct_{1,2}} = 0.07^{+0.07}_{-0.07}$ and $\Var{\ct_{1,2}} = 0.31^{+0.02}_{-0.07}$
in comparison to the full posterior which yields
$\mean{\ct_{1,2}} = 0.08^{+0.07}_{-0.07}$ and $\Var{\ct_{1,2}} = 0.28^{+0.05}_{-0.07}$.
All values quoted are medians and 90\% credible levels. 
Turning to the bottom panel of Fig.~\ref{fig:default-ppds}, the restricted posterior population distributions almost exactly reproduce the full posterior $a_{1,2}$ distributions.

The restricted posterior population distribution in $\ct_{1,2}$ is narrower
than the full posterior.
This is because the restricted posterior
is closer to the prior,
since filtering the catalog
data through effective spins neglects some information.
We find $\lnb = 3.3 \pm 0.63$ in favor of the \moddef component spin model over its
approximate pushforward.
Discrepancies between the restricted and full posteriors
may reflect features of the component spin pushforward
not captured by the parametric fits; cf. Fig.~\ref{fig:skew-kurt}.

Results obtained with the \modbiv model are shown in Fig.~\ref{fig:biv-ppds-joint}.
We recover support for a peak in the tilt distribution
at $(\ct_1, \ct_2) \sim (0.25, 0.25)$
which is reproduced
even when we restrict the population likelihood
to the marginal effective spin distributions.
We also find support for a negative correlation between primary and secondary tilts,
again reproduced in the restricted posterior.
Similarly,
there is support for a positive correlation between $a_1$--$a_2$
in the full and restricted posteriors.
The restricted analysis also reproduces the inferred preference for slow spins,
although with larger uncertainties.
With $\lnb = 1.5 \pm 0.80$,
we cannot prove that the \modbiv spin model
is favored over its approximate pushforward.

\begin{figure}[t]
    \centering
    \includegraphics[width=0.8\linewidth]{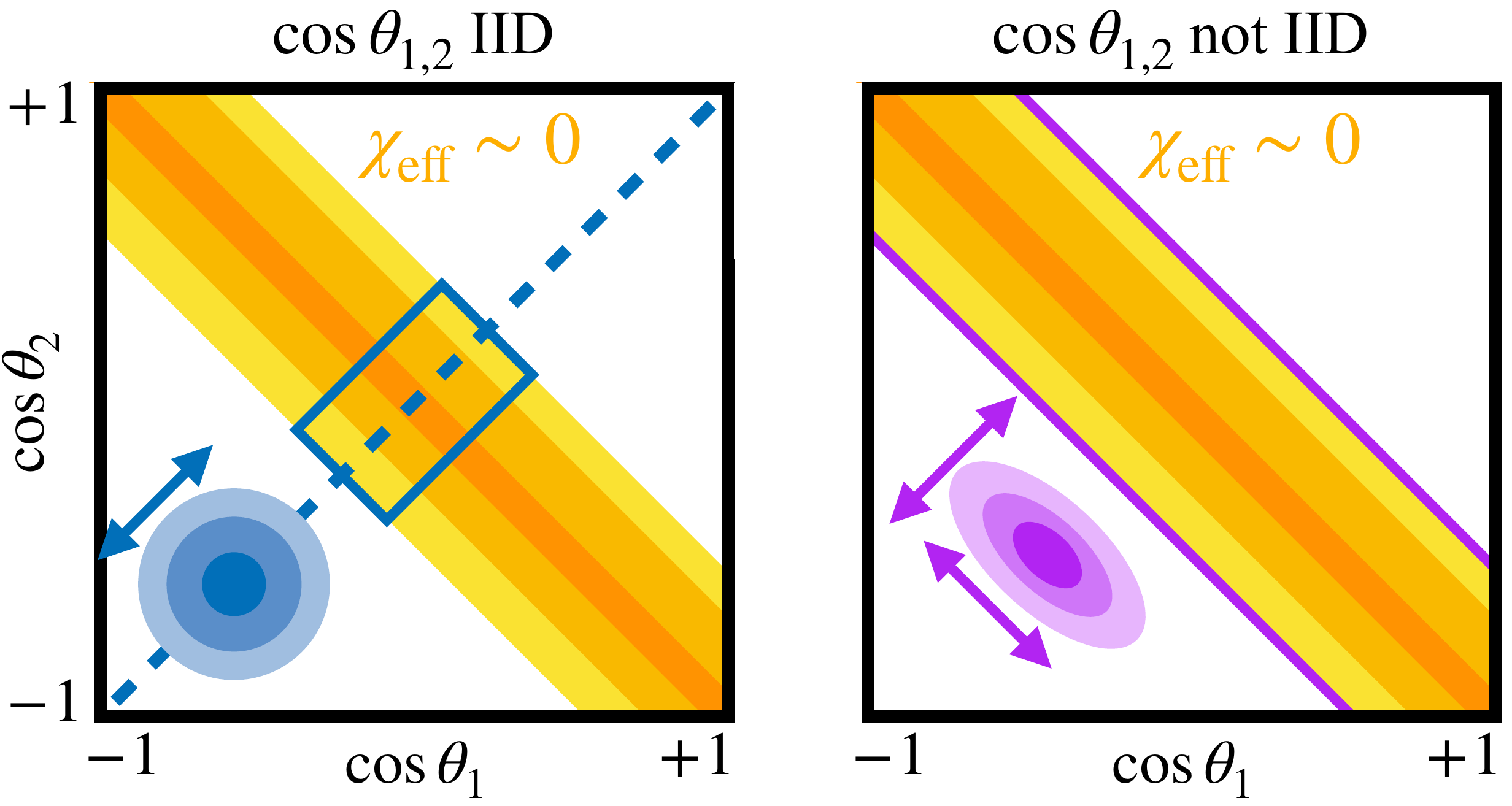}
    \caption{
    Cartoon projection of a zero-centered \chie spin population with finite width (yellow)
    onto the space of spin tilts,
    assuming equal \ac{BH} masses and equal, non-zero spin magnitudes;
    $\ct_2$ vs. $\ct_1$ slope remains negative
    even if its particular value changes with with $q$ and $a_{1,2}$.
    (\textit{Left}) Schematic of a model which assumes $\ct_{1,2}$ are \ac{IID},
    highlighting (box) where it intersects the \chie distribution.
    (\textit{Right}) Model which relaxes the \ac{IID} assumption.
    }
    \label{fig:toy}
\end{figure}

\textit{A Geometric Explanation}---Consider the limit where we perfectly measure the \chie distribution to be a delta function centered at zero.
Then, all draws from the component spin and mass ratio distribution must satisfy $a_1 \ct_1 + q \, a_2 \ct_2 = 0$, cf. Eq.~\eqref{eq:chieff-def}. 
\red{
The \moddef model
assumes a factorized distribution
$p(\ct_{1,2})p(a_{1,2}, q)$
and that $\ct_1$ and $\ct_2$ are \ac{NID}.
Under these assumptions (``\moddef-like''),
primary and secondary tilts are drawn without reference to how they will be paired
nor the magnitudes and mass ratio they will be paired with.
One way to ensure that $\chie = 0$ for every draw
is for the spin tilts to always be zero.
In the Supplemental Material,
we show that \moddef-like models can \textit{only}
match a zero-\chie population
at $(\ct_1, \ct_2) = 0$
or when $a_1 = q\,a_2$,
with $(a_1, a_2) = (0, 0)$
if we further assume magnitudes and mass ratio are independent.
}

If the \chie population has some finite width
then \red{a \moddef-like} model will infer that the tilt distribution
lies within an area that depends on the measured width of the \chie distribution.
In the left panel of Fig.~\ref{fig:toy}
we diagram the finite width case;
%in the left panel of Fig.~\ref{fig:toy};
a box indicates the overlap between \red{\ac{IID}} tilt modeling assumptions
and the inferred \chie population.
In the right panel of Fig.~\ref{fig:toy},
we relax the \ac{IID} assumption;
then, the tilt distribution can live anywhere
within the area set by our knowledge of the \chie population.
Inference with a model which relaxes the \ac{IID} assumption for primary and secondary tilts will find support for negatively correlated tilts.
This case corresponds to the \modbiv model.
Whereas in the \moddef model we found a peak near 
$\ct_{1,2} \sim 0$,
under the \modbiv model we found that
this feature broadens into a negative $\ct_1$--$\ct_2$ correlation
since the model can satisfy the \chie constraint surface at more than one point.

Ref.~\cite{LIGOScientific:2026ctl} finds that the \chie distribution is not centered at zero,
instead peaking slightly above zero
or with measurable skew.
The toy model in Fig.~\ref{fig:toy} extends to \chie distributions centered away from zero.
When the mean \chie increases (decreases)
the projection of the \chie population into spin tilts moves up and right (down and left).
Thus, the \chie population inferred by Ref.~\cite{LIGOScientific:2026ctl}
coincides with tilt distributions peaking at $\ct_{1,2} \gtrsim 0$,
as we found with both the \moddef and \modbiv models.

Realistically, we do not perfectly know the spin magnitude distributions either.
In general, 
measurements of the $\chie$--$\chip$--$q$ distribution
project to a constraint volume in $a_1$--$a_2$--$\ct_1$--$\ct_2$
with non-trivial geometry.
Many analyses assume that primary and secondary component spins
are identically distributed \red{(though see, e.g., Refs.~\cite{Mould:2022xeu, Adamcewicz:2023szp, Hussain:2024qzl} which relax this assumption).}
All, to our knowledge, neglect to jointly model the full space of component spins.
These prior assumptions take slices
of the constraint volume,
leading to spurious features in the component spin distribution
like peaks or correlations.

\textit{Outlook}---Recent advances in flexible population modeling could enable inference of joint models of the four component spin degrees of freedom \cite{Ray:2026qer}.
However, there are additional computational challenges
that come with high-dimensional population modeling \cite{Alvarez-Lopez:2025ltt}.
\red{
With a simulation study, Ref.~\cite{Miller:2024sui} found that population models which factorize the distribution of spin tilts, magnitudes, and mass ratio can qualitatively distinguish different component spin populations that yield identical \chie distributions.
However, their simulated populations also differed in \chip, which changed how accurately those models recovered the tilt distribution.
}
Our results suggest that analyses of current \ac{GW} catalogs
which make similarly strong assumptions about the component spin distribution
need only infer the $\red{\chib}$ distribution.
Additionally, our method can be extended to characterize the origin of spin inferences in e.g. mass--spin subpopulations, although that analysis is beyond the scope of this work.
We have shown that the apparent peak in the tilt distribution is spurious
because it can be understood as
the geometric point
where standard spin population modeling assumptions
intersect current constraints on the effective spin distribution.

\textit{Acknowledgements}:
We thank Sof\'ia \'Alvarez-L\'opez,
Maximiliano Isi, Konstantin Leyde, Matthew Mould, Cailin Plunkett, Mike Zevin, and Aaron Zimmerman  for discussions.
\red{We also thank Matthew Mould for internal review.}
N.E.W. and J.H. are supported by the National Science Foundation Graduate Research Fellowship Program under grant No. 2141064.
S.V. is partially supported by the NSF grant No. PHY-2045740.
The Flatiron Institute is a division of the Simons Foundation.
The authors are grateful for computational resources provided by
the LIGO Laboratory supported by National Science Foundation Grants PHY-0757058 and PHY-0823459,
and by subMIT at MIT Physics.
This material is based upon work supported by NSF's LIGO Laboratory which is a major facility fully funded by the National Science Foundation and has made use of data or software obtained from the Gravitational Wave Open Science Center (gwosc.org), a service of the LIGO Scientific Collaboration, the Virgo Collaboration, and KAGRA.

\bibliography{draft}

\begin{thebibliography}{99}%
\makeatletter
\providecommand \@ifxundefined [1]{%
 \@ifx{#1\undefined}
}%
\providecommand \@ifnum [1]{%
 \ifnum #1\expandafter \@firstoftwo
 \else \expandafter \@secondoftwo
 \fi
}%
\providecommand \@ifx [1]{%
 \ifx #1\expandafter \@firstoftwo
 \else \expandafter \@secondoftwo
 \fi
}%
\providecommand \natexlab [1]{#1}%
\providecommand \enquote  [1]{``#1''}%
\providecommand \bibnamefont  [1]{#1}%
\providecommand \bibfnamefont [1]{#1}%
\providecommand \citenamefont [1]{#1}%
\providecommand \href@noop [0]{\@secondoftwo}%
\providecommand \href [0]{\begingroup \@sanitize@url \@href}%
\providecommand \@href[1]{\@@startlink{#1}\@@href}%
\providecommand \@@href[1]{\endgroup#1\@@endlink}%
\providecommand \@sanitize@url [0]{\catcode `\\12\catcode `\$12\catcode
  `\&12\catcode `\#12\catcode `\^12\catcode `\_12\catcode `\%12\relax}%
\providecommand \@@startlink[1]{}%
\providecommand \@@endlink[0]{}%
\providecommand \url  [0]{\begingroup\@sanitize@url \@url }%
\providecommand \@url [1]{\endgroup\@href {#1}{\urlprefix }}%
\providecommand \urlprefix  [0]{URL }%
\providecommand \Eprint [0]{\href }%
\providecommand \doibase [0]{https://doi.org/}%
\providecommand \selectlanguage [0]{\@gobble}%
\providecommand \bibinfo  [0]{\@secondoftwo}%
\providecommand \bibfield  [0]{\@secondoftwo}%
\providecommand \translation [1]{[#1]}%
\providecommand \BibitemOpen [0]{}%
\providecommand \bibitemStop [0]{}%
\providecommand \bibitemNoStop [0]{.\EOS\space}%
\providecommand \EOS [0]{\spacefactor3000\relax}%
\providecommand \BibitemShut  [1]{\csname bibitem#1\endcsname}%
\let\auto@bib@innerbib\@empty
%</preamble>
\bibitem [{\citenamefont {Aasi}\ \emph {et~al.}(2015)\citenamefont {Aasi} \emph
  {et~al.}}]{LIGOScientific:2014pky}%
  \BibitemOpen
  \bibfield  {author} {\bibinfo {author} {\bibfnamefont {J.}~\bibnamefont
  {Aasi}} \emph {et~al.} (\bibinfo {collaboration} {LIGO Scientific}),\ }\href
  {https://doi.org/10.1088/0264-9381/32/7/074001} {\bibfield  {journal}
  {\bibinfo  {journal} {Class. Quant. Grav.}\ }\textbf {\bibinfo {volume}
  {32}},\ \bibinfo {pages} {074001} (\bibinfo {year} {2015})},\ \Eprint
  {https://arxiv.org/abs/1411.4547} {arXiv:1411.4547 [gr-qc]} \BibitemShut
  {NoStop}%
\bibitem [{\citenamefont {Acernese}\ \emph {et~al.}(2015)\citenamefont
  {Acernese} \emph {et~al.}}]{VIRGO:2014yos}%
  \BibitemOpen
  \bibfield  {author} {\bibinfo {author} {\bibfnamefont {F.}~\bibnamefont
  {Acernese}} \emph {et~al.} (\bibinfo {collaboration} {VIRGO}),\ }\href
  {https://doi.org/10.1088/0264-9381/32/2/024001} {\bibfield  {journal}
  {\bibinfo  {journal} {Class. Quant. Grav.}\ }\textbf {\bibinfo {volume}
  {32}},\ \bibinfo {pages} {024001} (\bibinfo {year} {2015})},\ \Eprint
  {https://arxiv.org/abs/1408.3978} {arXiv:1408.3978 [gr-qc]} \BibitemShut
  {NoStop}%
\bibitem [{\citenamefont {Akutsu}\ \emph {et~al.}(2021)\citenamefont {Akutsu}
  \emph {et~al.}}]{KAGRA:2020tym}%
  \BibitemOpen
  \bibfield  {author} {\bibinfo {author} {\bibfnamefont {T.}~\bibnamefont
  {Akutsu}} \emph {et~al.} (\bibinfo {collaboration} {KAGRA}),\ }\href
  {https://doi.org/10.1093/ptep/ptaa125} {\bibfield  {journal} {\bibinfo
  {journal} {PTEP}\ }\textbf {\bibinfo {volume} {2021}},\ \bibinfo {pages}
  {05A101} (\bibinfo {year} {2021})},\ \Eprint
  {https://arxiv.org/abs/2005.05574} {arXiv:2005.05574 [physics.ins-det]}
  \BibitemShut {NoStop}%
\bibitem [{\citenamefont {Abac}\ \emph
  {et~al.}(2026{\natexlab{a}})\citenamefont {Abac} \emph
  {et~al.}}]{LIGOScientific:2026sit}%
  \BibitemOpen
  \bibfield  {author} {\bibinfo {author} {\bibfnamefont {N.}~\bibnamefont
  {Abac}} \emph {et~al.} (\bibinfo {collaboration} {LIGO Scientific, VIRGO,
  KAGRA}),\ }\href@noop {} {\  (\bibinfo {year} {2026}{\natexlab{a}})},\
  \Eprint {https://arxiv.org/abs/2605.27223} {arXiv:2605.27223 [gr-qc]}
  \BibitemShut {NoStop}%
\bibitem [{\citenamefont {Abac}\ \emph
  {et~al.}(2026{\natexlab{b}})\citenamefont {Abac} \emph
  {et~al.}}]{LIGOScientific:2026wfs}%
  \BibitemOpen
  \bibfield  {author} {\bibinfo {author} {\bibfnamefont {A.~G.}\ \bibnamefont
  {Abac}} \emph {et~al.} (\bibinfo {collaboration} {LIGO Scientific, VIRGO,
  KAGRA}),\ }\href@noop {} {\  (\bibinfo {year} {2026}{\natexlab{b}})},\
  \Eprint {https://arxiv.org/abs/2605.27225} {arXiv:2605.27225 [gr-qc]}
  \BibitemShut {NoStop}%
\bibitem [{\citenamefont {Berti}\ and\ \citenamefont
  {Volonteri}(2008)}]{Berti:2008af}%
  \BibitemOpen
  \bibfield  {author} {\bibinfo {author} {\bibfnamefont {E.}~\bibnamefont
  {Berti}}\ and\ \bibinfo {author} {\bibfnamefont {M.}~\bibnamefont
  {Volonteri}},\ }\href {https://doi.org/10.1086/590379} {\bibfield  {journal}
  {\bibinfo  {journal} {Astrophys. J.}\ }\textbf {\bibinfo {volume} {684}},\
  \bibinfo {pages} {822} (\bibinfo {year} {2008})},\ \Eprint
  {https://arxiv.org/abs/0802.0025} {arXiv:0802.0025 [astro-ph]} \BibitemShut
  {NoStop}%
\bibitem [{\citenamefont {Gerosa}\ and\ \citenamefont
  {Fishbach}(2021)}]{Gerosa:2021mno}%
  \BibitemOpen
  \bibfield  {author} {\bibinfo {author} {\bibfnamefont {D.}~\bibnamefont
  {Gerosa}}\ and\ \bibinfo {author} {\bibfnamefont {M.}~\bibnamefont
  {Fishbach}},\ }\href {https://doi.org/10.1038/s41550-021-01398-w} {\bibfield
  {journal} {\bibinfo  {journal} {Nature Astron.}\ }\textbf {\bibinfo {volume}
  {5}},\ \bibinfo {pages} {749} (\bibinfo {year} {2021})},\ \Eprint
  {https://arxiv.org/abs/2105.03439} {arXiv:2105.03439 [astro-ph.HE]}
  \BibitemShut {NoStop}%
\bibitem [{\citenamefont {{Rodriguez}}\ \emph {et~al.}(2016)\citenamefont
  {{Rodriguez}}, \citenamefont {{Zevin}}, \citenamefont {{Pankow}},
  \citenamefont {{Kalogera}},\ and\ \citenamefont
  {{Rasio}}}]{2016ApJ...832L...2R}%
  \BibitemOpen
  \bibfield  {author} {\bibinfo {author} {\bibfnamefont {C.~L.}\ \bibnamefont
  {{Rodriguez}}}, \bibinfo {author} {\bibfnamefont {M.}~\bibnamefont
  {{Zevin}}}, \bibinfo {author} {\bibfnamefont {C.}~\bibnamefont {{Pankow}}},
  \bibinfo {author} {\bibfnamefont {V.}~\bibnamefont {{Kalogera}}},\ and\
  \bibinfo {author} {\bibfnamefont {F.~A.}\ \bibnamefont {{Rasio}}},\ }\href
  {https://doi.org/10.3847/2041-8205/832/1/L2} {\bibfield  {journal} {\bibinfo
  {journal} {Astrophys. J. Lett.}\ }\textbf {\bibinfo {volume} {832}},\
  \bibinfo {eid} {L2} (\bibinfo {year} {2016})},\ \Eprint
  {https://arxiv.org/abs/1609.05916} {arXiv:1609.05916 [astro-ph.HE]}
  \BibitemShut {NoStop}%
\bibitem [{\citenamefont {{Rodriguez}}\ \emph {et~al.}(2022)\citenamefont
  {{Rodriguez}}, \citenamefont {{Weatherford}}, \citenamefont {{Coughlin}},
  \citenamefont {{Amaro-Seoane}}, \citenamefont {{Breivik}}, \citenamefont
  {{Chatterjee}}, \citenamefont {{Fragione}}, \citenamefont
  {{K{\i}ro{\u{g}}lu}}, \citenamefont {{Kremer}}, \citenamefont {{Rui}},
  \citenamefont {{Ye}}, \citenamefont {{Zevin}},\ and\ \citenamefont
  {{Rasio}}}]{2022ApJS..258...22R}%
  \BibitemOpen
  \bibfield  {author} {\bibinfo {author} {\bibfnamefont {C.~L.}\ \bibnamefont
  {{Rodriguez}}}, \bibinfo {author} {\bibfnamefont {N.~C.}\ \bibnamefont
  {{Weatherford}}}, \bibinfo {author} {\bibfnamefont {S.~C.}\ \bibnamefont
  {{Coughlin}}}, \bibinfo {author} {\bibfnamefont {P.}~\bibnamefont
  {{Amaro-Seoane}}}, \bibinfo {author} {\bibfnamefont {K.}~\bibnamefont
  {{Breivik}}}, \bibinfo {author} {\bibfnamefont {S.}~\bibnamefont
  {{Chatterjee}}}, \bibinfo {author} {\bibfnamefont {G.}~\bibnamefont
  {{Fragione}}}, \bibinfo {author} {\bibfnamefont {F.}~\bibnamefont
  {{K{\i}ro{\u{g}}lu}}}, \bibinfo {author} {\bibfnamefont {K.}~\bibnamefont
  {{Kremer}}}, \bibinfo {author} {\bibfnamefont {N.~Z.}\ \bibnamefont {{Rui}}},
  \bibinfo {author} {\bibfnamefont {C.~S.}\ \bibnamefont {{Ye}}}, \bibinfo
  {author} {\bibfnamefont {M.}~\bibnamefont {{Zevin}}},\ and\ \bibinfo {author}
  {\bibfnamefont {F.~A.}\ \bibnamefont {{Rasio}}},\ }\href
  {https://doi.org/10.3847/1538-4365/ac2edf} {\bibfield  {journal} {\bibinfo
  {journal} {Astrophys. J. Supp.}\ }\textbf {\bibinfo {volume} {258}},\
  \bibinfo {eid} {22} (\bibinfo {year} {2022})},\ \Eprint
  {https://arxiv.org/abs/2106.02643} {arXiv:2106.02643 [astro-ph.GA]}
  \BibitemShut {NoStop}%
\bibitem [{\citenamefont {Kalogera}(2000)}]{Kalogera:1999tq}%
  \BibitemOpen
  \bibfield  {author} {\bibinfo {author} {\bibfnamefont {V.}~\bibnamefont
  {Kalogera}},\ }\href {https://doi.org/10.1086/309400} {\bibfield  {journal}
  {\bibinfo  {journal} {Astrophys. J.}\ }\textbf {\bibinfo {volume} {541}},\
  \bibinfo {pages} {319} (\bibinfo {year} {2000})},\ \Eprint
  {https://arxiv.org/abs/astro-ph/9911417} {arXiv:astro-ph/9911417}
  \BibitemShut {NoStop}%
\bibitem [{\citenamefont {Gerosa}\ \emph {et~al.}(2018)\citenamefont {Gerosa},
  \citenamefont {Berti}, \citenamefont {O'Shaughnessy}, \citenamefont
  {Belczynski}, \citenamefont {Kesden}, \citenamefont {Wysocki},\ and\
  \citenamefont {Gladysz}}]{Gerosa:2018wbw}%
  \BibitemOpen
  \bibfield  {author} {\bibinfo {author} {\bibfnamefont {D.}~\bibnamefont
  {Gerosa}}, \bibinfo {author} {\bibfnamefont {E.}~\bibnamefont {Berti}},
  \bibinfo {author} {\bibfnamefont {R.}~\bibnamefont {O'Shaughnessy}}, \bibinfo
  {author} {\bibfnamefont {K.}~\bibnamefont {Belczynski}}, \bibinfo {author}
  {\bibfnamefont {M.}~\bibnamefont {Kesden}}, \bibinfo {author} {\bibfnamefont
  {D.}~\bibnamefont {Wysocki}},\ and\ \bibinfo {author} {\bibfnamefont
  {W.}~\bibnamefont {Gladysz}},\ }\href
  {https://doi.org/10.1103/PhysRevD.98.084036} {\bibfield  {journal} {\bibinfo
  {journal} {Phys. Rev. D}\ }\textbf {\bibinfo {volume} {98}},\ \bibinfo
  {pages} {084036} (\bibinfo {year} {2018})},\ \Eprint
  {https://arxiv.org/abs/1808.02491} {arXiv:1808.02491 [astro-ph.HE]}
  \BibitemShut {NoStop}%
\bibitem [{\citenamefont {Belczynski}\ \emph {et~al.}(2020)\citenamefont
  {Belczynski} \emph {et~al.}}]{Belczynski:2017gds}%
  \BibitemOpen
  \bibfield  {author} {\bibinfo {author} {\bibfnamefont {K.}~\bibnamefont
  {Belczynski}} \emph {et~al.},\ }\href
  {https://doi.org/10.1051/0004-6361/201936528} {\bibfield  {journal} {\bibinfo
   {journal} {Astron. Astrophys.}\ }\textbf {\bibinfo {volume} {636}},\
  \bibinfo {pages} {A104} (\bibinfo {year} {2020})},\ \Eprint
  {https://arxiv.org/abs/1706.07053} {arXiv:1706.07053 [astro-ph.HE]}
  \BibitemShut {NoStop}%
\bibitem [{\citenamefont {Fuller}\ and\ \citenamefont
  {Ma}(2019)}]{Fuller:2019sxi}%
  \BibitemOpen
  \bibfield  {author} {\bibinfo {author} {\bibfnamefont {J.}~\bibnamefont
  {Fuller}}\ and\ \bibinfo {author} {\bibfnamefont {L.}~\bibnamefont {Ma}},\
  }\href {https://doi.org/10.3847/2041-8213/ab339b} {\bibfield  {journal}
  {\bibinfo  {journal} {Astrophys. J. Lett.}\ }\textbf {\bibinfo {volume}
  {881}},\ \bibinfo {pages} {L1} (\bibinfo {year} {2019})},\ \Eprint
  {https://arxiv.org/abs/1907.03714} {arXiv:1907.03714 [astro-ph.SR]}
  \BibitemShut {NoStop}%
\bibitem [{\citenamefont {Godfrey}\ \emph {et~al.}(2023)\citenamefont
  {Godfrey}, \citenamefont {Edelman},\ and\ \citenamefont
  {Farr}}]{Godfrey:2023oxb}%
  \BibitemOpen
  \bibfield  {author} {\bibinfo {author} {\bibfnamefont {J.}~\bibnamefont
  {Godfrey}}, \bibinfo {author} {\bibfnamefont {B.}~\bibnamefont {Edelman}},\
  and\ \bibinfo {author} {\bibfnamefont {B.}~\bibnamefont {Farr}},\ }\href@noop
  {} {\  (\bibinfo {year} {2023})},\ \Eprint {https://arxiv.org/abs/2304.01288}
  {arXiv:2304.01288 [astro-ph.HE]} \BibitemShut {NoStop}%
\bibitem [{\citenamefont {Li}\ \emph {et~al.}(2024)\citenamefont {Li},
  \citenamefont {Wang}, \citenamefont {Tang},\ and\ \citenamefont
  {Fan}}]{Li:2023yyt}%
  \BibitemOpen
  \bibfield  {author} {\bibinfo {author} {\bibfnamefont {Y.-J.}\ \bibnamefont
  {Li}}, \bibinfo {author} {\bibfnamefont {Y.-Z.}\ \bibnamefont {Wang}},
  \bibinfo {author} {\bibfnamefont {S.-P.}\ \bibnamefont {Tang}},\ and\
  \bibinfo {author} {\bibfnamefont {Y.-Z.}\ \bibnamefont {Fan}},\ }\href
  {https://doi.org/10.1103/PhysRevLett.133.051401} {\bibfield  {journal}
  {\bibinfo  {journal} {Phys. Rev. Lett.}\ }\textbf {\bibinfo {volume} {133}},\
  \bibinfo {pages} {051401} (\bibinfo {year} {2024})},\ \Eprint
  {https://arxiv.org/abs/2303.02973} {arXiv:2303.02973 [astro-ph.HE]}
  \BibitemShut {NoStop}%
\bibitem [{\citenamefont {Hussain}\ \emph {et~al.}(2026)\citenamefont
  {Hussain}, \citenamefont {Isi},\ and\ \citenamefont
  {Zimmerman}}]{Hussain:2024qzl}%
  \BibitemOpen
  \bibfield  {author} {\bibinfo {author} {\bibfnamefont {A.}~\bibnamefont
  {Hussain}}, \bibinfo {author} {\bibfnamefont {M.}~\bibnamefont {Isi}},\ and\
  \bibinfo {author} {\bibfnamefont {A.}~\bibnamefont {Zimmerman}},\ }\href
  {https://doi.org/10.3847/1538-4357/ae1574} {\bibfield  {journal} {\bibinfo
  {journal} {Astrophys. J.}\ }\textbf {\bibinfo {volume} {996}},\ \bibinfo
  {pages} {71} (\bibinfo {year} {2026})},\ \Eprint
  {https://arxiv.org/abs/2411.02252} {arXiv:2411.02252 [astro-ph.HE]}
  \BibitemShut {NoStop}%
\bibitem [{\citenamefont {Banagiri}\ \emph {et~al.}(2026)\citenamefont
  {Banagiri}, \citenamefont {Thrane},\ and\ \citenamefont
  {Lasky}}]{Banagiri:2025dmy}%
  \BibitemOpen
  \bibfield  {author} {\bibinfo {author} {\bibfnamefont {S.}~\bibnamefont
  {Banagiri}}, \bibinfo {author} {\bibfnamefont {E.}~\bibnamefont {Thrane}},\
  and\ \bibinfo {author} {\bibfnamefont {P.~D.}\ \bibnamefont {Lasky}},\ }\href
  {https://doi.org/10.1103/blyb-lqv6} {\bibfield  {journal} {\bibinfo
  {journal} {Phys. Rev. Lett.}\ }\textbf {\bibinfo {volume} {137}},\ \bibinfo
  {pages} {021403} (\bibinfo {year} {2026})},\ \Eprint
  {https://arxiv.org/abs/2509.15646} {arXiv:2509.15646 [astro-ph.HE]}
  \BibitemShut {NoStop}%
\bibitem [{\citenamefont {Farah}\ \emph {et~al.}(2026)\citenamefont {Farah},
  \citenamefont {Vijaykumar},\ and\ \citenamefont {Fishbach}}]{Farah:2026jlc}%
  \BibitemOpen
  \bibfield  {author} {\bibinfo {author} {\bibfnamefont {A.~M.}\ \bibnamefont
  {Farah}}, \bibinfo {author} {\bibfnamefont {A.}~\bibnamefont {Vijaykumar}},\
  and\ \bibinfo {author} {\bibfnamefont {M.}~\bibnamefont {Fishbach}},\ }\href
  {https://doi.org/10.3847/2041-8213/ae4e19} {\bibfield  {journal} {\bibinfo
  {journal} {Astrophys. J. Lett.}\ }\textbf {\bibinfo {volume} {1001}},\
  \bibinfo {pages} {L40} (\bibinfo {year} {2026})},\ \Eprint
  {https://arxiv.org/abs/2601.03456} {arXiv:2601.03456 [astro-ph.HE]}
  \BibitemShut {NoStop}%
\bibitem [{\citenamefont {Wolfe}\ \emph {et~al.}(2026)\citenamefont {Wolfe},
  \citenamefont {Vitale},\ and\ \citenamefont {Zevin}}]{Wolfe:2026meb}%
  \BibitemOpen
  \bibfield  {author} {\bibinfo {author} {\bibfnamefont {N.~E.}\ \bibnamefont
  {Wolfe}}, \bibinfo {author} {\bibfnamefont {S.}~\bibnamefont {Vitale}},\ and\
  \bibinfo {author} {\bibfnamefont {M.}~\bibnamefont {Zevin}} (\bibinfo
  {collaboration} {(Society of Physicists Interested in Non-aligned Spins,
  SPINS)}),\ }\href {https://doi.org/10.3847/2041-8213/ae90ae} {\bibfield
  {journal} {\bibinfo  {journal} {Astrophys. J. Lett.}\ }\textbf {\bibinfo
  {volume} {1007}},\ \bibinfo {pages} {L45} (\bibinfo {year} {2026})},\ \Eprint
  {https://arxiv.org/abs/2605.05300} {arXiv:2605.05300 [astro-ph.HE]}
  \BibitemShut {NoStop}%
\bibitem [{\citenamefont {Galaudage}(2026)}]{Galaudage:2026opk}%
  \BibitemOpen
  \bibfield  {author} {\bibinfo {author} {\bibfnamefont {S.}~\bibnamefont
  {Galaudage}},\ }\href@noop {} {\  (\bibinfo {year} {2026})},\ \Eprint
  {https://arxiv.org/abs/2605.25994} {arXiv:2605.25994 [astro-ph.HE]}
  \BibitemShut {NoStop}%
\bibitem [{\citenamefont {Guttman}\ \emph
  {et~al.}(2026{\natexlab{a}})\citenamefont {Guttman}, \citenamefont {Lasky},\
  and\ \citenamefont {Thrane}}]{Guttman:2026cnv}%
  \BibitemOpen
  \bibfield  {author} {\bibinfo {author} {\bibfnamefont {N.}~\bibnamefont
  {Guttman}}, \bibinfo {author} {\bibfnamefont {P.~D.}\ \bibnamefont {Lasky}},\
  and\ \bibinfo {author} {\bibfnamefont {E.}~\bibnamefont {Thrane}},\
  }\href@noop {} {\  (\bibinfo {year} {2026}{\natexlab{a}})},\ \Eprint
  {https://arxiv.org/abs/2607.22011} {arXiv:2607.22011 [astro-ph.HE]}
  \BibitemShut {NoStop}%
\bibitem [{\citenamefont {Golomb}\ and\ \citenamefont
  {Talbot}(2023)}]{Golomb:2022bon}%
  \BibitemOpen
  \bibfield  {author} {\bibinfo {author} {\bibfnamefont {J.}~\bibnamefont
  {Golomb}}\ and\ \bibinfo {author} {\bibfnamefont {C.}~\bibnamefont
  {Talbot}},\ }\href {https://doi.org/10.1103/PhysRevD.108.103009} {\bibfield
  {journal} {\bibinfo  {journal} {Phys. Rev. D}\ }\textbf {\bibinfo {volume}
  {108}},\ \bibinfo {pages} {103009} (\bibinfo {year} {2023})},\ \Eprint
  {https://arxiv.org/abs/2210.12287} {arXiv:2210.12287 [astro-ph.HE]}
  \BibitemShut {NoStop}%
\bibitem [{\citenamefont {Callister}\ and\ \citenamefont
  {Farr}(2024)}]{Callister:2023tgi}%
  \BibitemOpen
  \bibfield  {author} {\bibinfo {author} {\bibfnamefont {T.~A.}\ \bibnamefont
  {Callister}}\ and\ \bibinfo {author} {\bibfnamefont {W.~M.}\ \bibnamefont
  {Farr}},\ }\href {https://doi.org/10.1103/PhysRevX.14.021005} {\bibfield
  {journal} {\bibinfo  {journal} {Phys. Rev. X}\ }\textbf {\bibinfo {volume}
  {14}},\ \bibinfo {pages} {021005} (\bibinfo {year} {2024})},\ \Eprint
  {https://arxiv.org/abs/2302.07289} {arXiv:2302.07289 [astro-ph.HE]}
  \BibitemShut {NoStop}%
\bibitem [{\citenamefont {Vitale}\ \emph {et~al.}(2022)\citenamefont {Vitale},
  \citenamefont {Biscoveanu},\ and\ \citenamefont {Talbot}}]{Vitale:2022dpa}%
  \BibitemOpen
  \bibfield  {author} {\bibinfo {author} {\bibfnamefont {S.}~\bibnamefont
  {Vitale}}, \bibinfo {author} {\bibfnamefont {S.}~\bibnamefont {Biscoveanu}},\
  and\ \bibinfo {author} {\bibfnamefont {C.}~\bibnamefont {Talbot}},\ }\href
  {https://doi.org/10.1051/0004-6361/202245084} {\bibfield  {journal} {\bibinfo
   {journal} {Astron. Astrophys.}\ }\textbf {\bibinfo {volume} {668}},\
  \bibinfo {pages} {L2} (\bibinfo {year} {2022})},\ \Eprint
  {https://arxiv.org/abs/2209.06978} {arXiv:2209.06978 [astro-ph.HE]}
  \BibitemShut {NoStop}%
\bibitem [{\citenamefont {Stegmann}\ \emph {et~al.}(2026)\citenamefont
  {Stegmann}, \citenamefont {Antonini}, \citenamefont {Olejak}, \citenamefont
  {Biscoveanu}, \citenamefont {Raymond}, \citenamefont {Rinaldi},\ and\
  \citenamefont {Flanagan}}]{Stegmann:2025zkb}%
  \BibitemOpen
  \bibfield  {author} {\bibinfo {author} {\bibfnamefont {J.}~\bibnamefont
  {Stegmann}}, \bibinfo {author} {\bibfnamefont {F.}~\bibnamefont {Antonini}},
  \bibinfo {author} {\bibfnamefont {A.}~\bibnamefont {Olejak}}, \bibinfo
  {author} {\bibfnamefont {S.}~\bibnamefont {Biscoveanu}}, \bibinfo {author}
  {\bibfnamefont {V.}~\bibnamefont {Raymond}}, \bibinfo {author} {\bibfnamefont
  {S.}~\bibnamefont {Rinaldi}},\ and\ \bibinfo {author} {\bibfnamefont
  {E.}~\bibnamefont {Flanagan}},\ }\href
  {https://doi.org/10.3847/2041-8213/ae52ec} {\bibfield  {journal} {\bibinfo
  {journal} {Astrophys. J. Lett.}\ }\textbf {\bibinfo {volume} {1000}},\
  \bibinfo {pages} {L59} (\bibinfo {year} {2026})},\ \Eprint
  {https://arxiv.org/abs/2512.15873} {arXiv:2512.15873 [astro-ph.HE]}
  \BibitemShut {NoStop}%
\bibitem [{\citenamefont {Li}\ \emph {et~al.}(2026)\citenamefont {Li},
  \citenamefont {Wang}, \citenamefont {Tang},\ and\ \citenamefont
  {Fan}}]{Li:2025iux}%
  \BibitemOpen
  \bibfield  {author} {\bibinfo {author} {\bibfnamefont {Y.-J.}\ \bibnamefont
  {Li}}, \bibinfo {author} {\bibfnamefont {Y.-Z.}\ \bibnamefont {Wang}},
  \bibinfo {author} {\bibfnamefont {S.-P.}\ \bibnamefont {Tang}},\ and\
  \bibinfo {author} {\bibfnamefont {Y.-Z.}\ \bibnamefont {Fan}},\ }\href
  {https://doi.org/10.1103/qyfv-wkv1} {\bibfield  {journal} {\bibinfo
  {journal} {Phys. Rev. Lett.}\ }\textbf {\bibinfo {volume} {137}},\ \bibinfo
  {pages} {021407} (\bibinfo {year} {2026})},\ \Eprint
  {https://arxiv.org/abs/2509.23897} {arXiv:2509.23897 [astro-ph.HE]}
  \BibitemShut {NoStop}%
\bibitem [{\citenamefont {Abac}\ \emph
  {et~al.}(2026{\natexlab{c}})\citenamefont {Abac} \emph
  {et~al.}}]{LIGOScientific:2025pvj}%
  \BibitemOpen
  \bibfield  {author} {\bibinfo {author} {\bibfnamefont {A.~G.}\ \bibnamefont
  {Abac}} \emph {et~al.} (\bibinfo {collaboration} {LIGO Scientific, VIRGO,
  Virgo,, KAGRA}),\ }\href {https://doi.org/10.3847/2041-8213/ae771e}
  {\bibfield  {journal} {\bibinfo  {journal} {Astrophys. J. Lett.}\ }\textbf
  {\bibinfo {volume} {1005}},\ \bibinfo {pages} {L51} (\bibinfo {year}
  {2026}{\natexlab{c}})},\ \Eprint {https://arxiv.org/abs/2508.18083}
  {arXiv:2508.18083 [astro-ph.HE]} \BibitemShut {NoStop}%
\bibitem [{\citenamefont {Abac}\ \emph
  {et~al.}(2026{\natexlab{d}})\citenamefont {Abac} \emph
  {et~al.}}]{LIGOScientific:2026ctl}%
  \BibitemOpen
  \bibfield  {author} {\bibinfo {author} {\bibfnamefont {A.~G.}\ \bibnamefont
  {Abac}} \emph {et~al.} (\bibinfo {collaboration} {LIGO Scientific, VIRGO,
  KAGRA}),\ }\href@noop {} {\  (\bibinfo {year} {2026}{\natexlab{d}})},\
  \Eprint {https://arxiv.org/abs/2605.27226} {arXiv:2605.27226 [astro-ph.HE]}
  \BibitemShut {NoStop}%
\bibitem [{\citenamefont {Guttman}\ \emph
  {et~al.}(2026{\natexlab{b}})\citenamefont {Guttman}, \citenamefont {Payne},
  \citenamefont {Lasky},\ and\ \citenamefont {Thrane}}]{Guttman:2025jkv}%
  \BibitemOpen
  \bibfield  {author} {\bibinfo {author} {\bibfnamefont {N.}~\bibnamefont
  {Guttman}}, \bibinfo {author} {\bibfnamefont {E.}~\bibnamefont {Payne}},
  \bibinfo {author} {\bibfnamefont {P.~D.}\ \bibnamefont {Lasky}},\ and\
  \bibinfo {author} {\bibfnamefont {E.}~\bibnamefont {Thrane}},\ }\href
  {https://doi.org/10.3847/1538-4357/ae17af} {\bibfield  {journal} {\bibinfo
  {journal} {Astrophys. J.}\ }\textbf {\bibinfo {volume} {996}},\ \bibinfo
  {pages} {144} (\bibinfo {year} {2026}{\natexlab{b}})},\ \Eprint
  {https://arxiv.org/abs/2509.09876} {arXiv:2509.09876 [astro-ph.HE]}
  \BibitemShut {NoStop}%
\bibitem [{\citenamefont {Antonini}\ \emph {et~al.}(2018)\citenamefont
  {Antonini}, \citenamefont {Rodriguez}, \citenamefont {Petrovich},\ and\
  \citenamefont {Fischer}}]{Antonini:2017tgo}%
  \BibitemOpen
  \bibfield  {author} {\bibinfo {author} {\bibfnamefont {F.}~\bibnamefont
  {Antonini}}, \bibinfo {author} {\bibfnamefont {C.~L.}\ \bibnamefont
  {Rodriguez}}, \bibinfo {author} {\bibfnamefont {C.}~\bibnamefont
  {Petrovich}},\ and\ \bibinfo {author} {\bibfnamefont {C.~L.}\ \bibnamefont
  {Fischer}},\ }\href {https://doi.org/10.1093/mnrasl/sly126} {\bibfield
  {journal} {\bibinfo  {journal} {Mon. Not. Roy. Astron. Soc.}\ }\textbf
  {\bibinfo {volume} {480}},\ \bibinfo {pages} {L58} (\bibinfo {year}
  {2018})},\ \Eprint {https://arxiv.org/abs/1711.07142} {arXiv:1711.07142
  [astro-ph.HE]} \BibitemShut {NoStop}%
\bibitem [{\citenamefont {Rodriguez}\ and\ \citenamefont
  {Antonini}(2018)}]{Rodriguez:2018jqu}%
  \BibitemOpen
  \bibfield  {author} {\bibinfo {author} {\bibfnamefont {C.~L.}\ \bibnamefont
  {Rodriguez}}\ and\ \bibinfo {author} {\bibfnamefont {F.}~\bibnamefont
  {Antonini}},\ }\href {https://doi.org/10.3847/1538-4357/aacea4} {\bibfield
  {journal} {\bibinfo  {journal} {Astrophys. J.}\ }\textbf {\bibinfo {volume}
  {863}},\ \bibinfo {pages} {7} (\bibinfo {year} {2018})},\ \Eprint
  {https://arxiv.org/abs/1805.08212} {arXiv:1805.08212 [astro-ph.HE]}
  \BibitemShut {NoStop}%
\bibitem [{\citenamefont {Liu}\ and\ \citenamefont {Lai}(2018)}]{Liu:2018nrf}%
  \BibitemOpen
  \bibfield  {author} {\bibinfo {author} {\bibfnamefont {B.}~\bibnamefont
  {Liu}}\ and\ \bibinfo {author} {\bibfnamefont {D.}~\bibnamefont {Lai}},\
  }\href {https://doi.org/10.3847/1538-4357/aad09f} {\bibfield  {journal}
  {\bibinfo  {journal} {Astrophys. J.}\ }\textbf {\bibinfo {volume} {863}},\
  \bibinfo {pages} {68} (\bibinfo {year} {2018})},\ \Eprint
  {https://arxiv.org/abs/1805.03202} {arXiv:1805.03202 [astro-ph.HE]}
  \BibitemShut {NoStop}%
\bibitem [{\citenamefont {Yu}\ \emph {et~al.}(2020)\citenamefont {Yu},
  \citenamefont {Ma}, \citenamefont {Giesler},\ and\ \citenamefont
  {Chen}}]{Yu:2020iqj}%
  \BibitemOpen
  \bibfield  {author} {\bibinfo {author} {\bibfnamefont {H.}~\bibnamefont
  {Yu}}, \bibinfo {author} {\bibfnamefont {S.}~\bibnamefont {Ma}}, \bibinfo
  {author} {\bibfnamefont {M.}~\bibnamefont {Giesler}},\ and\ \bibinfo {author}
  {\bibfnamefont {Y.}~\bibnamefont {Chen}},\ }\href
  {https://doi.org/10.1103/PhysRevD.102.123009} {\bibfield  {journal} {\bibinfo
   {journal} {Phys. Rev. D}\ }\textbf {\bibinfo {volume} {102}},\ \bibinfo
  {pages} {123009} (\bibinfo {year} {2020})},\ \Eprint
  {https://arxiv.org/abs/2007.12978} {arXiv:2007.12978 [gr-qc]} \BibitemShut
  {NoStop}%
\bibitem [{\citenamefont {Su}\ \emph {et~al.}(2021)\citenamefont {Su},
  \citenamefont {Lai},\ and\ \citenamefont {Liu}}]{Su:2020vda}%
  \BibitemOpen
  \bibfield  {author} {\bibinfo {author} {\bibfnamefont {Y.}~\bibnamefont
  {Su}}, \bibinfo {author} {\bibfnamefont {D.}~\bibnamefont {Lai}},\ and\
  \bibinfo {author} {\bibfnamefont {B.}~\bibnamefont {Liu}},\ }\href
  {https://doi.org/10.1103/PhysRevD.103.063040} {\bibfield  {journal} {\bibinfo
   {journal} {Phys. Rev. D}\ }\textbf {\bibinfo {volume} {103}},\ \bibinfo
  {pages} {063040} (\bibinfo {year} {2021})},\ \Eprint
  {https://arxiv.org/abs/2010.11951} {arXiv:2010.11951 [gr-qc]} \BibitemShut
  {NoStop}%
\bibitem [{\citenamefont {Biscoveanu}(2026)}]{Biscoveanu:2026ikx}%
  \BibitemOpen
  \bibfield  {author} {\bibinfo {author} {\bibfnamefont {S.}~\bibnamefont
  {Biscoveanu}},\ }\href@noop {} {\  (\bibinfo {year} {2026})},\ \Eprint
  {https://arxiv.org/abs/2606.06209} {arXiv:2606.06209 [gr-qc]} \BibitemShut
  {NoStop}%
\bibitem [{\citenamefont {Vitale}\ and\ \citenamefont
  {Mould}(2025)}]{Vitale:2025lms}%
  \BibitemOpen
  \bibfield  {author} {\bibinfo {author} {\bibfnamefont {S.}~\bibnamefont
  {Vitale}}\ and\ \bibinfo {author} {\bibfnamefont {M.}~\bibnamefont {Mould}}
  (\bibinfo {collaboration} {Society of Physicists Interested in Non-aligned
  Spins, SPINS, (Society of Physicists Interested in Non-aligned Spins,
  SPINS){\textdagger}}),\ }\href {https://doi.org/10.1103/drsl-n3wz} {\bibfield
   {journal} {\bibinfo  {journal} {Phys. Rev. D}\ }\textbf {\bibinfo {volume}
  {112}},\ \bibinfo {pages} {083015} (\bibinfo {year} {2025})},\ \Eprint
  {https://arxiv.org/abs/2505.14875} {arXiv:2505.14875 [astro-ph.HE]}
  \BibitemShut {NoStop}%
\bibitem [{\citenamefont {Corelli}\ \emph {et~al.}(2026)\citenamefont
  {Corelli}, \citenamefont {Gerosa}, \citenamefont {Mould},\ and\ \citenamefont
  {Fabbri}}]{Corelli:2026thw}%
  \BibitemOpen
  \bibfield  {author} {\bibinfo {author} {\bibfnamefont {A.}~\bibnamefont
  {Corelli}}, \bibinfo {author} {\bibfnamefont {D.}~\bibnamefont {Gerosa}},
  \bibinfo {author} {\bibfnamefont {M.}~\bibnamefont {Mould}},\ and\ \bibinfo
  {author} {\bibfnamefont {C.~M.}\ \bibnamefont {Fabbri}},\ }\href@noop {} {\
  (\bibinfo {year} {2026})},\ \Eprint {https://arxiv.org/abs/2603.00239}
  {arXiv:2603.00239 [astro-ph.HE]} \BibitemShut {NoStop}%
\bibitem [{\citenamefont {Abbott}\ \emph {et~al.}(2023)\citenamefont {Abbott}
  \emph {et~al.}}]{KAGRA:2021duu}%
  \BibitemOpen
  \bibfield  {author} {\bibinfo {author} {\bibfnamefont {R.}~\bibnamefont
  {Abbott}} \emph {et~al.} (\bibinfo {collaboration} {KAGRA, VIRGO, LIGO
  Scientific}),\ }\href {https://doi.org/10.1103/PhysRevX.13.011048} {\bibfield
   {journal} {\bibinfo  {journal} {Phys. Rev. X}\ }\textbf {\bibinfo {volume}
  {13}},\ \bibinfo {pages} {011048} (\bibinfo {year} {2023})},\ \Eprint
  {https://arxiv.org/abs/2111.03634} {arXiv:2111.03634 [astro-ph.HE]}
  \BibitemShut {NoStop}%
\bibitem [{\citenamefont {van~der Sluys}\ \emph {et~al.}(2008)\citenamefont
  {van~der Sluys}, \citenamefont {R{\"o}ver}, \citenamefont {Stroeer},
  \citenamefont {Raymond}, \citenamefont {Mandel}, \citenamefont {Christensen},
  \citenamefont {Kalogera}, \citenamefont {Meyer},\ and\ \citenamefont
  {Vecchio}}]{vanderSluys:2007st}%
  \BibitemOpen
  \bibfield  {author} {\bibinfo {author} {\bibfnamefont {M.~V.}\ \bibnamefont
  {van~der Sluys}}, \bibinfo {author} {\bibfnamefont {C.}~\bibnamefont
  {R{\"o}ver}}, \bibinfo {author} {\bibfnamefont {A.}~\bibnamefont {Stroeer}},
  \bibinfo {author} {\bibfnamefont {V.}~\bibnamefont {Raymond}}, \bibinfo
  {author} {\bibfnamefont {I.}~\bibnamefont {Mandel}}, \bibinfo {author}
  {\bibfnamefont {N.}~\bibnamefont {Christensen}}, \bibinfo {author}
  {\bibfnamefont {V.}~\bibnamefont {Kalogera}}, \bibinfo {author}
  {\bibfnamefont {R.}~\bibnamefont {Meyer}},\ and\ \bibinfo {author}
  {\bibfnamefont {A.}~\bibnamefont {Vecchio}},\ }\href
  {https://doi.org/10.1086/595279} {\bibfield  {journal} {\bibinfo  {journal}
  {Astrophys. J. Lett.}\ }\textbf {\bibinfo {volume} {688}},\ \bibinfo {pages}
  {L61} (\bibinfo {year} {2008})},\ \Eprint {https://arxiv.org/abs/0710.1897}
  {arXiv:0710.1897 [astro-ph]} \BibitemShut {NoStop}%
\bibitem [{\citenamefont {van~der Sluys}\ \emph {et~al.}(2009)\citenamefont
  {van~der Sluys}, \citenamefont {Mandel}, \citenamefont {Raymond},
  \citenamefont {Kalogera}, \citenamefont {Rover},\ and\ \citenamefont
  {Christensen}}]{vanderSluys:2009bf}%
  \BibitemOpen
  \bibfield  {author} {\bibinfo {author} {\bibfnamefont {M.}~\bibnamefont
  {van~der Sluys}}, \bibinfo {author} {\bibfnamefont {I.}~\bibnamefont
  {Mandel}}, \bibinfo {author} {\bibfnamefont {V.}~\bibnamefont {Raymond}},
  \bibinfo {author} {\bibfnamefont {V.}~\bibnamefont {Kalogera}}, \bibinfo
  {author} {\bibfnamefont {C.}~\bibnamefont {Rover}},\ and\ \bibinfo {author}
  {\bibfnamefont {N.}~\bibnamefont {Christensen}},\ }\href
  {https://doi.org/10.1088/0264-9381/26/20/204010} {\bibfield  {journal}
  {\bibinfo  {journal} {Class. Quant. Grav.}\ }\textbf {\bibinfo {volume}
  {26}},\ \bibinfo {pages} {204010} (\bibinfo {year} {2009})},\ \Eprint
  {https://arxiv.org/abs/0905.1323} {arXiv:0905.1323 [gr-qc]} \BibitemShut
  {NoStop}%
\bibitem [{\citenamefont {Pankow}\ \emph {et~al.}(2017)\citenamefont {Pankow},
  \citenamefont {Sampson}, \citenamefont {Perri}, \citenamefont {Chase},
  \citenamefont {Coughlin}, \citenamefont {Zevin},\ and\ \citenamefont
  {Kalogera}}]{Pankow:2016udj}%
  \BibitemOpen
  \bibfield  {author} {\bibinfo {author} {\bibfnamefont {C.}~\bibnamefont
  {Pankow}}, \bibinfo {author} {\bibfnamefont {L.}~\bibnamefont {Sampson}},
  \bibinfo {author} {\bibfnamefont {L.}~\bibnamefont {Perri}}, \bibinfo
  {author} {\bibfnamefont {E.}~\bibnamefont {Chase}}, \bibinfo {author}
  {\bibfnamefont {S.}~\bibnamefont {Coughlin}}, \bibinfo {author}
  {\bibfnamefont {M.}~\bibnamefont {Zevin}},\ and\ \bibinfo {author}
  {\bibfnamefont {V.}~\bibnamefont {Kalogera}},\ }\href
  {https://doi.org/10.3847/1538-4357/834/2/154} {\bibfield  {journal} {\bibinfo
   {journal} {Astrophys. J.}\ }\textbf {\bibinfo {volume} {834}},\ \bibinfo
  {pages} {154} (\bibinfo {year} {2017})},\ \Eprint
  {https://arxiv.org/abs/1610.05633} {arXiv:1610.05633 [astro-ph.HE]}
  \BibitemShut {NoStop}%
\bibitem [{\citenamefont {Vitale}\ \emph
  {et~al.}(2017{\natexlab{a}})\citenamefont {Vitale}, \citenamefont {Lynch},
  \citenamefont {Raymond}, \citenamefont {Sturani}, \citenamefont {Veitch},\
  and\ \citenamefont {Graff}}]{Vitale:2016avz}%
  \BibitemOpen
  \bibfield  {author} {\bibinfo {author} {\bibfnamefont {S.}~\bibnamefont
  {Vitale}}, \bibinfo {author} {\bibfnamefont {R.}~\bibnamefont {Lynch}},
  \bibinfo {author} {\bibfnamefont {V.}~\bibnamefont {Raymond}}, \bibinfo
  {author} {\bibfnamefont {R.}~\bibnamefont {Sturani}}, \bibinfo {author}
  {\bibfnamefont {J.}~\bibnamefont {Veitch}},\ and\ \bibinfo {author}
  {\bibfnamefont {P.}~\bibnamefont {Graff}},\ }\href
  {https://doi.org/10.1103/PhysRevD.95.064053} {\bibfield  {journal} {\bibinfo
  {journal} {Phys. Rev. D}\ }\textbf {\bibinfo {volume} {95}},\ \bibinfo
  {pages} {064053} (\bibinfo {year} {2017}{\natexlab{a}})},\ \Eprint
  {https://arxiv.org/abs/1611.01122} {arXiv:1611.01122 [gr-qc]} \BibitemShut
  {NoStop}%
\bibitem [{\citenamefont {P{\"u}rrer}\ \emph {et~al.}(2016)\citenamefont
  {P{\"u}rrer}, \citenamefont {Hannam},\ and\ \citenamefont
  {Ohme}}]{Purrer:2015nkh}%
  \BibitemOpen
  \bibfield  {author} {\bibinfo {author} {\bibfnamefont {M.}~\bibnamefont
  {P{\"u}rrer}}, \bibinfo {author} {\bibfnamefont {M.}~\bibnamefont {Hannam}},\
  and\ \bibinfo {author} {\bibfnamefont {F.}~\bibnamefont {Ohme}},\ }\href
  {https://doi.org/10.1103/PhysRevD.93.084042} {\bibfield  {journal} {\bibinfo
  {journal} {Phys. Rev. D}\ }\textbf {\bibinfo {volume} {93}},\ \bibinfo
  {pages} {084042} (\bibinfo {year} {2016})},\ \Eprint
  {https://arxiv.org/abs/1512.04955} {arXiv:1512.04955 [gr-qc]} \BibitemShut
  {NoStop}%
\bibitem [{\citenamefont {Shaik}\ \emph {et~al.}(2020)\citenamefont {Shaik},
  \citenamefont {Lange}, \citenamefont {Field}, \citenamefont {O'Shaughnessy},
  \citenamefont {Varma}, \citenamefont {Kidder}, \citenamefont {Pfeiffer},\
  and\ \citenamefont {Wysocki}}]{Shaik:2019dym}%
  \BibitemOpen
  \bibfield  {author} {\bibinfo {author} {\bibfnamefont {F.~H.}\ \bibnamefont
  {Shaik}}, \bibinfo {author} {\bibfnamefont {J.}~\bibnamefont {Lange}},
  \bibinfo {author} {\bibfnamefont {S.~E.}\ \bibnamefont {Field}}, \bibinfo
  {author} {\bibfnamefont {R.}~\bibnamefont {O'Shaughnessy}}, \bibinfo {author}
  {\bibfnamefont {V.}~\bibnamefont {Varma}}, \bibinfo {author} {\bibfnamefont
  {L.~E.}\ \bibnamefont {Kidder}}, \bibinfo {author} {\bibfnamefont {H.~P.}\
  \bibnamefont {Pfeiffer}},\ and\ \bibinfo {author} {\bibfnamefont
  {D.}~\bibnamefont {Wysocki}},\ }\href
  {https://doi.org/10.1103/PhysRevD.101.124054} {\bibfield  {journal} {\bibinfo
   {journal} {Phys. Rev. D}\ }\textbf {\bibinfo {volume} {101}},\ \bibinfo
  {pages} {124054} (\bibinfo {year} {2020})},\ \Eprint
  {https://arxiv.org/abs/1911.02693} {arXiv:1911.02693 [gr-qc]} \BibitemShut
  {NoStop}%
\bibitem [{\citenamefont {Pratten}\ \emph {et~al.}(2020)\citenamefont
  {Pratten}, \citenamefont {Schmidt}, \citenamefont {Buscicchio},\ and\
  \citenamefont {Thomas}}]{Pratten:2020igi}%
  \BibitemOpen
  \bibfield  {author} {\bibinfo {author} {\bibfnamefont {G.}~\bibnamefont
  {Pratten}}, \bibinfo {author} {\bibfnamefont {P.}~\bibnamefont {Schmidt}},
  \bibinfo {author} {\bibfnamefont {R.}~\bibnamefont {Buscicchio}},\ and\
  \bibinfo {author} {\bibfnamefont {L.~M.}\ \bibnamefont {Thomas}},\ }\href
  {https://doi.org/10.1103/PhysRevResearch.2.043096} {\bibfield  {journal}
  {\bibinfo  {journal} {Phys. Rev. Res.}\ }\textbf {\bibinfo {volume} {2}},\
  \bibinfo {pages} {043096} (\bibinfo {year} {2020})},\ \Eprint
  {https://arxiv.org/abs/2006.16153} {arXiv:2006.16153 [gr-qc]} \BibitemShut
  {NoStop}%
\bibitem [{\citenamefont {Green}\ \emph {et~al.}(2021)\citenamefont {Green},
  \citenamefont {Hoy}, \citenamefont {Fairhurst}, \citenamefont {Hannam},
  \citenamefont {Pannarale},\ and\ \citenamefont {Thomas}}]{Green:2020ptm}%
  \BibitemOpen
  \bibfield  {author} {\bibinfo {author} {\bibfnamefont {R.}~\bibnamefont
  {Green}}, \bibinfo {author} {\bibfnamefont {C.}~\bibnamefont {Hoy}}, \bibinfo
  {author} {\bibfnamefont {S.}~\bibnamefont {Fairhurst}}, \bibinfo {author}
  {\bibfnamefont {M.}~\bibnamefont {Hannam}}, \bibinfo {author} {\bibfnamefont
  {F.}~\bibnamefont {Pannarale}},\ and\ \bibinfo {author} {\bibfnamefont
  {C.}~\bibnamefont {Thomas}},\ }\href
  {https://doi.org/10.1103/PhysRevD.103.124023} {\bibfield  {journal} {\bibinfo
   {journal} {Phys. Rev. D}\ }\textbf {\bibinfo {volume} {103}},\ \bibinfo
  {pages} {124023} (\bibinfo {year} {2021})},\ \Eprint
  {https://arxiv.org/abs/2010.04131} {arXiv:2010.04131 [gr-qc]} \BibitemShut
  {NoStop}%
\bibitem [{\citenamefont {Biscoveanu}\ \emph {et~al.}(2021)\citenamefont
  {Biscoveanu}, \citenamefont {Isi}, \citenamefont {Varma},\ and\ \citenamefont
  {Vitale}}]{Biscoveanu:2021nvg}%
  \BibitemOpen
  \bibfield  {author} {\bibinfo {author} {\bibfnamefont {S.}~\bibnamefont
  {Biscoveanu}}, \bibinfo {author} {\bibfnamefont {M.}~\bibnamefont {Isi}},
  \bibinfo {author} {\bibfnamefont {V.}~\bibnamefont {Varma}},\ and\ \bibinfo
  {author} {\bibfnamefont {S.}~\bibnamefont {Vitale}},\ }\href
  {https://doi.org/10.1103/PhysRevD.104.103018} {\bibfield  {journal} {\bibinfo
   {journal} {Phys. Rev. D}\ }\textbf {\bibinfo {volume} {104}},\ \bibinfo
  {pages} {103018} (\bibinfo {year} {2021})},\ \Eprint
  {https://arxiv.org/abs/2106.06492} {arXiv:2106.06492 [gr-qc]} \BibitemShut
  {NoStop}%
\bibitem [{\citenamefont {Krishnendu}\ and\ \citenamefont
  {Ohme}(2022)}]{Krishnendu:2021cyi}%
  \BibitemOpen
  \bibfield  {author} {\bibinfo {author} {\bibfnamefont {N.~V.}\ \bibnamefont
  {Krishnendu}}\ and\ \bibinfo {author} {\bibfnamefont {F.}~\bibnamefont
  {Ohme}},\ }\href {https://doi.org/10.1103/PhysRevD.105.064012} {\bibfield
  {journal} {\bibinfo  {journal} {Phys. Rev. D}\ }\textbf {\bibinfo {volume}
  {105}},\ \bibinfo {pages} {064012} (\bibinfo {year} {2022})},\ \Eprint
  {https://arxiv.org/abs/2110.00766} {arXiv:2110.00766 [gr-qc]} \BibitemShut
  {NoStop}%
\bibitem [{\citenamefont {Miller}\ \emph {et~al.}(2025)\citenamefont {Miller},
  \citenamefont {Isi}, \citenamefont {Chatziioannou}, \citenamefont {Varma},\
  and\ \citenamefont {Hourihane}}]{Miller:2025eak}%
  \BibitemOpen
  \bibfield  {author} {\bibinfo {author} {\bibfnamefont {S.~J.}\ \bibnamefont
  {Miller}}, \bibinfo {author} {\bibfnamefont {M.}~\bibnamefont {Isi}},
  \bibinfo {author} {\bibfnamefont {K.}~\bibnamefont {Chatziioannou}}, \bibinfo
  {author} {\bibfnamefont {V.}~\bibnamefont {Varma}},\ and\ \bibinfo {author}
  {\bibfnamefont {S.}~\bibnamefont {Hourihane}},\ }\href
  {https://doi.org/10.1103/xq5g-zm7z} {\bibfield  {journal} {\bibinfo
  {journal} {Phys. Rev. D}\ }\textbf {\bibinfo {volume} {112}},\ \bibinfo
  {pages} {104046} (\bibinfo {year} {2025})},\ \Eprint
  {https://arxiv.org/abs/2505.14573} {arXiv:2505.14573 [gr-qc]} \BibitemShut
  {NoStop}%
\bibitem [{Note1()}]{Note1}%
  \BibitemOpen
  \bibinfo {note} {See Refs.~\cite {Gerosa:2020aiw, Thomas:2020uqj} for
  extensions of \protect \ensuremath {\chi _\protect \mathrm {p}}\protect
  \xspace which more accurately capture \ac {GW} inspiral spin precession
  morphology.}\BibitemShut {Stop}%
\bibitem [{\citenamefont {Damour}(2001)}]{Damour:2001tu}%
  \BibitemOpen
  \bibfield  {author} {\bibinfo {author} {\bibfnamefont {T.}~\bibnamefont
  {Damour}},\ }\href {https://doi.org/10.1103/PhysRevD.64.124013} {\bibfield
  {journal} {\bibinfo  {journal} {Phys. Rev. D}\ }\textbf {\bibinfo {volume}
  {64}},\ \bibinfo {pages} {124013} (\bibinfo {year} {2001})},\ \Eprint
  {https://arxiv.org/abs/gr-qc/0103018} {arXiv:gr-qc/0103018} \BibitemShut
  {NoStop}%
\bibitem [{\citenamefont {Racine}(2008)}]{Racine:2008qv}%
  \BibitemOpen
  \bibfield  {author} {\bibinfo {author} {\bibfnamefont {E.}~\bibnamefont
  {Racine}},\ }\href {https://doi.org/10.1103/PhysRevD.78.044021} {\bibfield
  {journal} {\bibinfo  {journal} {Phys. Rev. D}\ }\textbf {\bibinfo {volume}
  {78}},\ \bibinfo {pages} {044021} (\bibinfo {year} {2008})},\ \Eprint
  {https://arxiv.org/abs/0803.1820} {arXiv:0803.1820 [gr-qc]} \BibitemShut
  {NoStop}%
\bibitem [{\citenamefont {{Blanchet}}(2014)}]{2014LRR....17....2B}%
  \BibitemOpen
  \bibfield  {author} {\bibinfo {author} {\bibfnamefont {L.}~\bibnamefont
  {{Blanchet}}},\ }\href {https://doi.org/10.12942/lrr-2014-2} {\bibfield
  {journal} {\bibinfo  {journal} {Living Reviews in Relativity}\ }\textbf
  {\bibinfo {volume} {17}},\ \bibinfo {eid} {2} (\bibinfo {year} {2014})},\
  \Eprint {https://arxiv.org/abs/1310.1528} {arXiv:1310.1528 [gr-qc]}
  \BibitemShut {NoStop}%
\bibitem [{\citenamefont {Schmidt}\ \emph {et~al.}(2015)\citenamefont
  {Schmidt}, \citenamefont {Ohme},\ and\ \citenamefont
  {Hannam}}]{Schmidt:2014iyl}%
  \BibitemOpen
  \bibfield  {author} {\bibinfo {author} {\bibfnamefont {P.}~\bibnamefont
  {Schmidt}}, \bibinfo {author} {\bibfnamefont {F.}~\bibnamefont {Ohme}},\ and\
  \bibinfo {author} {\bibfnamefont {M.}~\bibnamefont {Hannam}},\ }\href
  {https://doi.org/10.1103/PhysRevD.91.024043} {\bibfield  {journal} {\bibinfo
  {journal} {Phys. Rev. D}\ }\textbf {\bibinfo {volume} {91}},\ \bibinfo
  {pages} {024043} (\bibinfo {year} {2015})},\ \Eprint
  {https://arxiv.org/abs/1408.1810} {arXiv:1408.1810 [gr-qc]} \BibitemShut
  {NoStop}%
\bibitem [{\citenamefont {Plunkett}\ \emph {et~al.}(2026)\citenamefont
  {Plunkett} \emph {et~al.}}]{plunkett2026prep}%
  \BibitemOpen
  \bibfield  {author} {\bibinfo {author} {\bibfnamefont {C.}~\bibnamefont
  {Plunkett}} \emph {et~al.}} (\bibinfo {year} {2026}),\ \bibinfo {note}
  {manuscript in preparation}\BibitemShut {NoStop}%
\bibitem [{\citenamefont {Miller}\ \emph {et~al.}(2024)\citenamefont {Miller},
  \citenamefont {Ko}, \citenamefont {Callister},\ and\ \citenamefont
  {Chatziioannou}}]{Miller:2024sui}%
  \BibitemOpen
  \bibfield  {author} {\bibinfo {author} {\bibfnamefont {S.~J.}\ \bibnamefont
  {Miller}}, \bibinfo {author} {\bibfnamefont {Z.}~\bibnamefont {Ko}}, \bibinfo
  {author} {\bibfnamefont {T.}~\bibnamefont {Callister}},\ and\ \bibinfo
  {author} {\bibfnamefont {K.}~\bibnamefont {Chatziioannou}},\ }\href
  {https://doi.org/10.1103/PhysRevD.109.104036} {\bibfield  {journal} {\bibinfo
   {journal} {Phys. Rev. D}\ }\textbf {\bibinfo {volume} {109}},\ \bibinfo
  {pages} {104036} (\bibinfo {year} {2024})},\ \Eprint
  {https://arxiv.org/abs/2401.05613} {arXiv:2401.05613 [gr-qc]} \BibitemShut
  {NoStop}%
\bibitem [{\citenamefont {Loredo}\ and\ \citenamefont
  {Lamb}(2002)}]{Loredo:2001rx}%
  \BibitemOpen
  \bibfield  {author} {\bibinfo {author} {\bibfnamefont {T.~J.}\ \bibnamefont
  {Loredo}}\ and\ \bibinfo {author} {\bibfnamefont {D.~Q.}\ \bibnamefont
  {Lamb}},\ }\href {https://doi.org/10.1103/PhysRevD.65.063002} {\bibfield
  {journal} {\bibinfo  {journal} {Phys. Rev. D}\ }\textbf {\bibinfo {volume}
  {65}},\ \bibinfo {pages} {063002} (\bibinfo {year} {2002})},\ \Eprint
  {https://arxiv.org/abs/astro-ph/0107260} {arXiv:astro-ph/0107260}
  \BibitemShut {NoStop}%
\bibitem [{\citenamefont {{Mandel}}\ \emph {et~al.}(2019)\citenamefont
  {{Mandel}}, \citenamefont {{Farr}},\ and\ \citenamefont
  {{Gair}}}]{2019MNRAS.486.1086M}%
  \BibitemOpen
  \bibfield  {author} {\bibinfo {author} {\bibfnamefont {I.}~\bibnamefont
  {{Mandel}}}, \bibinfo {author} {\bibfnamefont {W.~M.}\ \bibnamefont
  {{Farr}}},\ and\ \bibinfo {author} {\bibfnamefont {J.~R.}\ \bibnamefont
  {{Gair}}},\ }\href {https://doi.org/10.1093/mnras/stz896} {\bibfield
  {journal} {\bibinfo  {journal} {Mon. Not. Roy. Astron. Soc.}\ }\textbf
  {\bibinfo {volume} {486}},\ \bibinfo {pages} {1086} (\bibinfo {year}
  {2019})},\ \Eprint {https://arxiv.org/abs/1809.02063} {arXiv:1809.02063
  [physics.data-an]} \BibitemShut {NoStop}%
\bibitem [{\citenamefont {Vitale}\ \emph {et~al.}(2020)\citenamefont {Vitale},
  \citenamefont {Gerosa}, \citenamefont {Farr},\ and\ \citenamefont
  {Taylor}}]{Vitale:2020aaz}%
  \BibitemOpen
  \bibfield  {author} {\bibinfo {author} {\bibfnamefont {S.}~\bibnamefont
  {Vitale}}, \bibinfo {author} {\bibfnamefont {D.}~\bibnamefont {Gerosa}},
  \bibinfo {author} {\bibfnamefont {W.~M.}\ \bibnamefont {Farr}},\ and\
  \bibinfo {author} {\bibfnamefont {S.~R.}\ \bibnamefont {Taylor}},\ }in\ \href
  {https://doi.org/10.1007/978-981-15-4702-7_45-1} {\emph {\bibinfo {booktitle}
  {Handbook of Gravitational Wave Astronomy}}},\ \bibinfo {editor} {edited by\
  \bibinfo {editor} {\bibfnamefont {C.}~\bibnamefont {Bambi}}, \bibinfo
  {editor} {\bibfnamefont {S.}~\bibnamefont {Katsanevas}},\ and\ \bibinfo
  {editor} {\bibfnamefont {K.~D.}\ \bibnamefont {Kokkotas}}}\ (\bibinfo
  {publisher} {Springer},\ \bibinfo {address} {Singapore},\ \bibinfo {year}
  {2020})\ pp.\ \bibinfo {pages} {1--60}\BibitemShut {NoStop}%
\bibitem [{\citenamefont {Tiwari}(2018)}]{Tiwari:2017ndi}%
  \BibitemOpen
  \bibfield  {author} {\bibinfo {author} {\bibfnamefont {V.}~\bibnamefont
  {Tiwari}},\ }\href {https://doi.org/10.1088/1361-6382/aac89d} {\bibfield
  {journal} {\bibinfo  {journal} {Class. Quant. Grav.}\ }\textbf {\bibinfo
  {volume} {35}},\ \bibinfo {pages} {145009} (\bibinfo {year} {2018})},\
  \Eprint {https://arxiv.org/abs/1712.00482} {arXiv:1712.00482 [astro-ph.HE]}
  \BibitemShut {NoStop}%
\bibitem [{\citenamefont {Essick}\ and\ \citenamefont
  {Farr}(2022)}]{Essick:2022ojx}%
  \BibitemOpen
  \bibfield  {author} {\bibinfo {author} {\bibfnamefont {R.}~\bibnamefont
  {Essick}}\ and\ \bibinfo {author} {\bibfnamefont {W.}~\bibnamefont {Farr}},\
  }\href@noop {} {\  (\bibinfo {year} {2022})},\ \Eprint
  {https://arxiv.org/abs/2204.00461} {arXiv:2204.00461 [astro-ph.IM]}
  \BibitemShut {NoStop}%
\bibitem [{\citenamefont {Talbot}\ and\ \citenamefont
  {Golomb}(2023)}]{Talbot:2023pex}%
  \BibitemOpen
  \bibfield  {author} {\bibinfo {author} {\bibfnamefont {C.}~\bibnamefont
  {Talbot}}\ and\ \bibinfo {author} {\bibfnamefont {J.}~\bibnamefont
  {Golomb}},\ }\href {https://doi.org/10.1093/mnras/stad2968} {\bibfield
  {journal} {\bibinfo  {journal} {Mon. Not. Roy. Astron. Soc.}\ }\textbf
  {\bibinfo {volume} {526}},\ \bibinfo {pages} {3495} (\bibinfo {year}
  {2023})},\ \Eprint {https://arxiv.org/abs/2304.06138} {arXiv:2304.06138
  [astro-ph.IM]} \BibitemShut {NoStop}%
\bibitem [{\citenamefont {Heinzel}\ and\ \citenamefont
  {Vitale}(2025)}]{Heinzel:2025ogf}%
  \BibitemOpen
  \bibfield  {author} {\bibinfo {author} {\bibfnamefont {J.}~\bibnamefont
  {Heinzel}}\ and\ \bibinfo {author} {\bibfnamefont {S.}~\bibnamefont
  {Vitale}},\ }\href@noop {} {\  (\bibinfo {year} {2025})},\ \Eprint
  {https://arxiv.org/abs/2509.07221} {arXiv:2509.07221 [astro-ph.HE]}
  \BibitemShut {NoStop}%
\bibitem [{\citenamefont {Gangardt}\ \emph {et~al.}(2022)\citenamefont
  {Gangardt}, \citenamefont {Gerosa}, \citenamefont {Kesden}, \citenamefont
  {De~Renzis},\ and\ \citenamefont {Steinle}}]{Gangardt:2022ltd}%
  \BibitemOpen
  \bibfield  {author} {\bibinfo {author} {\bibfnamefont {D.}~\bibnamefont
  {Gangardt}}, \bibinfo {author} {\bibfnamefont {D.}~\bibnamefont {Gerosa}},
  \bibinfo {author} {\bibfnamefont {M.}~\bibnamefont {Kesden}}, \bibinfo
  {author} {\bibfnamefont {V.}~\bibnamefont {De~Renzis}},\ and\ \bibinfo
  {author} {\bibfnamefont {N.}~\bibnamefont {Steinle}},\ }\href
  {https://doi.org/10.1103/PhysRevD.106.024019} {\bibfield  {journal} {\bibinfo
   {journal} {Phys. Rev. D}\ }\textbf {\bibinfo {volume} {106}},\ \bibinfo
  {pages} {024019} (\bibinfo {year} {2022})},\ \bibinfo {note} {[Erratum:
  Phys.Rev.D 107, 109901 (2023)]},\ \Eprint {https://arxiv.org/abs/2204.00026}
  {arXiv:2204.00026 [gr-qc]} \BibitemShut {NoStop}%
\bibitem [{\citenamefont {Essick}\ \emph {et~al.}(2025)\citenamefont {Essick}
  \emph {et~al.}}]{Essick:2025zed}%
  \BibitemOpen
  \bibfield  {author} {\bibinfo {author} {\bibfnamefont {R.}~\bibnamefont
  {Essick}} \emph {et~al.},\ }\href {https://doi.org/10.1103/44x3-hv3y}
  {\bibfield  {journal} {\bibinfo  {journal} {Phys. Rev. D}\ }\textbf {\bibinfo
  {volume} {112}},\ \bibinfo {pages} {102001} (\bibinfo {year} {2025})},\
  \Eprint {https://arxiv.org/abs/2508.10638} {arXiv:2508.10638 [gr-qc]}
  \BibitemShut {NoStop}%
\bibitem [{Note2()}]{Note2}%
  \BibitemOpen
  \bibinfo {note} {When $p(s \mid \lambda ) \propto 1$ we recover the Jacobian
  derived by Refs.~\cite {Callister:2021gxf, Iwaya:2024zzq}. See Ref.~\cite
  {Farr:2017uvj} for other analytic pushforwards.}\BibitemShut {Stop}%
\bibitem [{Note3()}]{Note3}%
  \BibitemOpen
  \bibinfo {note} {See similarly Ref.~\cite {Callister:2020vyz} which inferred
  the \protect \ensuremath {\chi _\protect \mathrm {eff}}\protect \xspace
  distribution with a population model estimated for a physically motivated \ac
  {MC} simulation of component spins.}\BibitemShut {Stop}%
\bibitem [{\citenamefont {Kass}\ and\ \citenamefont
  {Raftery}(1995)}]{kass1995bayes}%
  \BibitemOpen
  \bibfield  {author} {\bibinfo {author} {\bibfnamefont {R.~E.}\ \bibnamefont
  {Kass}}\ and\ \bibinfo {author} {\bibfnamefont {A.~E.}\ \bibnamefont
  {Raftery}},\ }\href@noop {} {\bibfield  {journal} {\bibinfo  {journal}
  {Journal of the american statistical association}\ }\textbf {\bibinfo
  {volume} {90}},\ \bibinfo {pages} {773} (\bibinfo {year} {1995})}\BibitemShut
  {NoStop}%
\bibitem [{\citenamefont {Vitale}\ \emph
  {et~al.}(2017{\natexlab{b}})\citenamefont {Vitale}, \citenamefont {Lynch},
  \citenamefont {Sturani},\ and\ \citenamefont {Graff}}]{Vitale:2015tea}%
  \BibitemOpen
  \bibfield  {author} {\bibinfo {author} {\bibfnamefont {S.}~\bibnamefont
  {Vitale}}, \bibinfo {author} {\bibfnamefont {R.}~\bibnamefont {Lynch}},
  \bibinfo {author} {\bibfnamefont {R.}~\bibnamefont {Sturani}},\ and\ \bibinfo
  {author} {\bibfnamefont {P.}~\bibnamefont {Graff}},\ }\href
  {https://doi.org/10.1088/1361-6382/aa552e} {\bibfield  {journal} {\bibinfo
  {journal} {Class. Quant. Grav.}\ }\textbf {\bibinfo {volume} {34}},\ \bibinfo
  {pages} {03LT01} (\bibinfo {year} {2017}{\natexlab{b}})},\ \Eprint
  {https://arxiv.org/abs/1503.04307} {arXiv:1503.04307 [gr-qc]} \BibitemShut
  {NoStop}%
\bibitem [{\citenamefont {Talbot}\ and\ \citenamefont
  {Thrane}(2017)}]{Talbot:2017yur}%
  \BibitemOpen
  \bibfield  {author} {\bibinfo {author} {\bibfnamefont {C.}~\bibnamefont
  {Talbot}}\ and\ \bibinfo {author} {\bibfnamefont {E.}~\bibnamefont
  {Thrane}},\ }\href {https://doi.org/10.1103/PhysRevD.96.023012} {\bibfield
  {journal} {\bibinfo  {journal} {Phys. Rev. D}\ }\textbf {\bibinfo {volume}
  {96}},\ \bibinfo {pages} {023012} (\bibinfo {year} {2017})},\ \Eprint
  {https://arxiv.org/abs/1704.08370} {arXiv:1704.08370 [astro-ph.HE]}
  \BibitemShut {NoStop}%
\bibitem [{\citenamefont {Mould}\ \emph {et~al.}(2022)\citenamefont {Mould},
  \citenamefont {Gerosa}, \citenamefont {Broekgaarden},\ and\ \citenamefont
  {Steinle}}]{Mould:2022xeu}%
  \BibitemOpen
  \bibfield  {author} {\bibinfo {author} {\bibfnamefont {M.}~\bibnamefont
  {Mould}}, \bibinfo {author} {\bibfnamefont {D.}~\bibnamefont {Gerosa}},
  \bibinfo {author} {\bibfnamefont {F.~S.}\ \bibnamefont {Broekgaarden}},\ and\
  \bibinfo {author} {\bibfnamefont {N.}~\bibnamefont {Steinle}},\ }\href
  {https://doi.org/10.1093/mnras/stac2859} {\bibfield  {journal} {\bibinfo
  {journal} {Mon. Not. Roy. Astron. Soc.}\ }\textbf {\bibinfo {volume} {517}},\
  \bibinfo {pages} {2738} (\bibinfo {year} {2022})},\ \Eprint
  {https://arxiv.org/abs/2205.12329} {arXiv:2205.12329 [astro-ph.HE]}
  \BibitemShut {NoStop}%
\bibitem [{\citenamefont {Adamcewicz}\ \emph {et~al.}(2024)\citenamefont
  {Adamcewicz}, \citenamefont {Galaudage}, \citenamefont {Lasky},\ and\
  \citenamefont {Thrane}}]{Adamcewicz:2023szp}%
  \BibitemOpen
  \bibfield  {author} {\bibinfo {author} {\bibfnamefont {C.}~\bibnamefont
  {Adamcewicz}}, \bibinfo {author} {\bibfnamefont {S.}~\bibnamefont
  {Galaudage}}, \bibinfo {author} {\bibfnamefont {P.~D.}\ \bibnamefont
  {Lasky}},\ and\ \bibinfo {author} {\bibfnamefont {E.}~\bibnamefont
  {Thrane}},\ }\href {https://doi.org/10.3847/2041-8213/ad2df2} {\bibfield
  {journal} {\bibinfo  {journal} {Astrophys. J. Lett.}\ }\textbf {\bibinfo
  {volume} {964}},\ \bibinfo {pages} {L6} (\bibinfo {year} {2024})},\ \Eprint
  {https://arxiv.org/abs/2311.05182} {arXiv:2311.05182 [astro-ph.HE]}
  \BibitemShut {NoStop}%
\bibitem [{\citenamefont {Ray}\ and\ \citenamefont
  {Kalogera}(2026)}]{Ray:2026qer}%
  \BibitemOpen
  \bibfield  {author} {\bibinfo {author} {\bibfnamefont {A.}~\bibnamefont
  {Ray}}\ and\ \bibinfo {author} {\bibfnamefont {V.}~\bibnamefont {Kalogera}},\
  }\href@noop {} {\  (\bibinfo {year} {2026})},\ \Eprint
  {https://arxiv.org/abs/2607.28622} {arXiv:2607.28622 [astro-ph.HE]}
  \BibitemShut {NoStop}%
\bibitem [{\citenamefont {Alvarez-Lopez}\ \emph {et~al.}(2026)\citenamefont
  {Alvarez-Lopez}, \citenamefont {Heinzel}, \citenamefont {Mould},\ and\
  \citenamefont {Vitale}}]{Alvarez-Lopez:2025ltt}%
  \BibitemOpen
  \bibfield  {author} {\bibinfo {author} {\bibfnamefont {S.}~\bibnamefont
  {Alvarez-Lopez}}, \bibinfo {author} {\bibfnamefont {J.}~\bibnamefont
  {Heinzel}}, \bibinfo {author} {\bibfnamefont {M.}~\bibnamefont {Mould}},\
  and\ \bibinfo {author} {\bibfnamefont {S.}~\bibnamefont {Vitale}},\ }\href
  {https://doi.org/10.3847/2041-8213/ae81a9} {\bibfield  {journal} {\bibinfo
  {journal} {Astrophys. J. Lett.}\ }\textbf {\bibinfo {volume} {1006}},\
  \bibinfo {pages} {L5} (\bibinfo {year} {2026})},\ \Eprint
  {https://arxiv.org/abs/2506.20731} {arXiv:2506.20731 [astro-ph.HE]}
  \BibitemShut {NoStop}%
\bibitem [{\citenamefont {Gerosa}\ \emph {et~al.}(2021)\citenamefont {Gerosa},
  \citenamefont {Mould}, \citenamefont {Gangardt}, \citenamefont {Schmidt},
  \citenamefont {Pratten},\ and\ \citenamefont {Thomas}}]{Gerosa:2020aiw}%
  \BibitemOpen
  \bibfield  {author} {\bibinfo {author} {\bibfnamefont {D.}~\bibnamefont
  {Gerosa}}, \bibinfo {author} {\bibfnamefont {M.}~\bibnamefont {Mould}},
  \bibinfo {author} {\bibfnamefont {D.}~\bibnamefont {Gangardt}}, \bibinfo
  {author} {\bibfnamefont {P.}~\bibnamefont {Schmidt}}, \bibinfo {author}
  {\bibfnamefont {G.}~\bibnamefont {Pratten}},\ and\ \bibinfo {author}
  {\bibfnamefont {L.~M.}\ \bibnamefont {Thomas}},\ }\href
  {https://doi.org/10.1103/PhysRevD.103.064067} {\bibfield  {journal} {\bibinfo
   {journal} {Phys. Rev. D}\ }\textbf {\bibinfo {volume} {103}},\ \bibinfo
  {pages} {064067} (\bibinfo {year} {2021})},\ \Eprint
  {https://arxiv.org/abs/2011.11948} {arXiv:2011.11948 [gr-qc]} \BibitemShut
  {NoStop}%
\bibitem [{\citenamefont {Thomas}\ \emph {et~al.}(2021)\citenamefont {Thomas},
  \citenamefont {Schmidt},\ and\ \citenamefont {Pratten}}]{Thomas:2020uqj}%
  \BibitemOpen
  \bibfield  {author} {\bibinfo {author} {\bibfnamefont {L.~M.}\ \bibnamefont
  {Thomas}}, \bibinfo {author} {\bibfnamefont {P.}~\bibnamefont {Schmidt}},\
  and\ \bibinfo {author} {\bibfnamefont {G.}~\bibnamefont {Pratten}},\ }\href
  {https://doi.org/10.1103/PhysRevD.103.083022} {\bibfield  {journal} {\bibinfo
   {journal} {Phys. Rev. D}\ }\textbf {\bibinfo {volume} {103}},\ \bibinfo
  {pages} {083022} (\bibinfo {year} {2021})},\ \Eprint
  {https://arxiv.org/abs/2012.02209} {arXiv:2012.02209 [gr-qc]} \BibitemShut
  {NoStop}%
\bibitem [{\citenamefont {Callister}(2021)}]{Callister:2021gxf}%
  \BibitemOpen
  \bibfield  {author} {\bibinfo {author} {\bibfnamefont {T.}~\bibnamefont
  {Callister}},\ }\href@noop {} {\  (\bibinfo {year} {2021})},\ \Eprint
  {https://arxiv.org/abs/2104.09508} {arXiv:2104.09508 [gr-qc]} \BibitemShut
  {NoStop}%
\bibitem [{\citenamefont {Iwaya}\ \emph {et~al.}(2025)\citenamefont {Iwaya},
  \citenamefont {Kobayashi}, \citenamefont {Morisaki}, \citenamefont
  {Hotokezaka},\ and\ \citenamefont {Kinugawa}}]{Iwaya:2024zzq}%
  \BibitemOpen
  \bibfield  {author} {\bibinfo {author} {\bibfnamefont {M.}~\bibnamefont
  {Iwaya}}, \bibinfo {author} {\bibfnamefont {K.}~\bibnamefont {Kobayashi}},
  \bibinfo {author} {\bibfnamefont {S.}~\bibnamefont {Morisaki}}, \bibinfo
  {author} {\bibfnamefont {K.}~\bibnamefont {Hotokezaka}},\ and\ \bibinfo
  {author} {\bibfnamefont {T.}~\bibnamefont {Kinugawa}},\ }\href
  {https://doi.org/10.1103/PhysRevD.111.103046} {\bibfield  {journal} {\bibinfo
   {journal} {Phys. Rev. D}\ }\textbf {\bibinfo {volume} {111}},\ \bibinfo
  {pages} {103046} (\bibinfo {year} {2025})},\ \Eprint
  {https://arxiv.org/abs/2412.14551} {arXiv:2412.14551 [gr-qc]} \BibitemShut
  {NoStop}%
\bibitem [{\citenamefont {Farr}\ \emph {et~al.}(2017)\citenamefont {Farr},
  \citenamefont {Stevenson}, \citenamefont {Coleman~Miller}, \citenamefont
  {Mandel}, \citenamefont {Farr},\ and\ \citenamefont
  {Vecchio}}]{Farr:2017uvj}%
  \BibitemOpen
  \bibfield  {author} {\bibinfo {author} {\bibfnamefont {W.~M.}\ \bibnamefont
  {Farr}}, \bibinfo {author} {\bibfnamefont {S.}~\bibnamefont {Stevenson}},
  \bibinfo {author} {\bibfnamefont {M.}~\bibnamefont {Coleman~Miller}},
  \bibinfo {author} {\bibfnamefont {I.}~\bibnamefont {Mandel}}, \bibinfo
  {author} {\bibfnamefont {B.}~\bibnamefont {Farr}},\ and\ \bibinfo {author}
  {\bibfnamefont {A.}~\bibnamefont {Vecchio}},\ }\href
  {https://doi.org/10.1038/nature23453} {\bibfield  {journal} {\bibinfo
  {journal} {Nature}\ }\textbf {\bibinfo {volume} {548}},\ \bibinfo {pages}
  {426} (\bibinfo {year} {2017})},\ \Eprint {https://arxiv.org/abs/1706.01385}
  {arXiv:1706.01385 [astro-ph.HE]} \BibitemShut {NoStop}%
\bibitem [{\citenamefont {Callister}\ \emph {et~al.}(2021)\citenamefont
  {Callister}, \citenamefont {Farr},\ and\ \citenamefont
  {Renzo}}]{Callister:2020vyz}%
  \BibitemOpen
  \bibfield  {author} {\bibinfo {author} {\bibfnamefont {T.~A.}\ \bibnamefont
  {Callister}}, \bibinfo {author} {\bibfnamefont {W.~M.}\ \bibnamefont
  {Farr}},\ and\ \bibinfo {author} {\bibfnamefont {M.}~\bibnamefont {Renzo}},\
  }\href {https://doi.org/10.3847/1538-4357/ac1347} {\bibfield  {journal}
  {\bibinfo  {journal} {Astrophys. J.}\ }\textbf {\bibinfo {volume} {920}},\
  \bibinfo {pages} {157} (\bibinfo {year} {2021})},\ \Eprint
  {https://arxiv.org/abs/2011.09570} {arXiv:2011.09570 [astro-ph.HE]}
  \BibitemShut {NoStop}%
\bibitem [{\citenamefont {Subbotin}(1923)}]{subbotin1923law}%
  \BibitemOpen
  \bibfield  {author} {\bibinfo {author} {\bibfnamefont {M.~T.}\ \bibnamefont
  {Subbotin}},\ }\href@noop {} {\bibfield  {journal} {\bibinfo  {journal}
  {Sbornik: Mathematics}\ }\textbf {\bibinfo {volume} {31}},\ \bibinfo {pages}
  {296} (\bibinfo {year} {1923})}\BibitemShut {NoStop}%
\bibitem [{\citenamefont {Kullback}\ and\ \citenamefont {Leibler}(1951)}]{kl}%
  \BibitemOpen
  \bibfield  {author} {\bibinfo {author} {\bibfnamefont {S.}~\bibnamefont
  {Kullback}}\ and\ \bibinfo {author} {\bibfnamefont {R.~A.}\ \bibnamefont
  {Leibler}},\ }\href {https://api.semanticscholar.org/CorpusID:120349231}
  {\bibfield  {journal} {\bibinfo  {journal} {Annals of Mathematical
  Statistics}\ }\textbf {\bibinfo {volume} {22}},\ \bibinfo {pages} {79}
  (\bibinfo {year} {1951})}\BibitemShut {NoStop}%
\bibitem [{\citenamefont {Rader}\ \emph {et~al.}(2024)\citenamefont {Rader},
  \citenamefont {Lyons},\ and\ \citenamefont {Kidger}}]{optimistix2024}%
  \BibitemOpen
  \bibfield  {author} {\bibinfo {author} {\bibfnamefont {J.}~\bibnamefont
  {Rader}}, \bibinfo {author} {\bibfnamefont {T.}~\bibnamefont {Lyons}},\ and\
  \bibinfo {author} {\bibfnamefont {P.}~\bibnamefont {Kidger}},\ }\href@noop {}
  {\bibfield  {journal} {\bibinfo  {journal} {arXiv:2402.09983}\ } (\bibinfo
  {year} {2024})}\BibitemShut {NoStop}%
\bibitem [{\citenamefont {Banagiri}\ \emph {et~al.}(2025)\citenamefont
  {Banagiri}, \citenamefont {Callister}, \citenamefont {Adamcewicz},
  \citenamefont {Doctor},\ and\ \citenamefont {Kalogera}}]{Banagiri:2025dxo}%
  \BibitemOpen
  \bibfield  {author} {\bibinfo {author} {\bibfnamefont {S.}~\bibnamefont
  {Banagiri}}, \bibinfo {author} {\bibfnamefont {T.~A.}\ \bibnamefont
  {Callister}}, \bibinfo {author} {\bibfnamefont {C.}~\bibnamefont
  {Adamcewicz}}, \bibinfo {author} {\bibfnamefont {Z.}~\bibnamefont {Doctor}},\
  and\ \bibinfo {author} {\bibfnamefont {V.}~\bibnamefont {Kalogera}},\ }\href
  {https://doi.org/10.3847/1538-4357/adf4c6} {\bibfield  {journal} {\bibinfo
  {journal} {Astrophys. J.}\ }\textbf {\bibinfo {volume} {990}},\ \bibinfo
  {pages} {147} (\bibinfo {year} {2025})},\ \Eprint
  {https://arxiv.org/abs/2501.06712} {arXiv:2501.06712 [astro-ph.HE]}
  \BibitemShut {NoStop}%
\bibitem [{\citenamefont {{Papamakarios}}\ \emph {et~al.}(2019)\citenamefont
  {{Papamakarios}}, \citenamefont {{Nalisnick}}, \citenamefont {{Jimenez
  Rezende}}, \citenamefont {{Mohamed}},\ and\ \citenamefont
  {{Lakshminarayanan}}}]{2019arXiv191202762P}%
  \BibitemOpen
  \bibfield  {author} {\bibinfo {author} {\bibfnamefont {G.}~\bibnamefont
  {{Papamakarios}}}, \bibinfo {author} {\bibfnamefont {E.}~\bibnamefont
  {{Nalisnick}}}, \bibinfo {author} {\bibfnamefont {D.}~\bibnamefont {{Jimenez
  Rezende}}}, \bibinfo {author} {\bibfnamefont {S.}~\bibnamefont {{Mohamed}}},\
  and\ \bibinfo {author} {\bibfnamefont {B.}~\bibnamefont
  {{Lakshminarayanan}}},\ }\href {https://doi.org/10.48550/arXiv.1912.02762}
  {\bibfield  {journal} {\bibinfo  {journal} {arXiv e-prints}\ ,\ \bibinfo
  {eid} {arXiv:1912.02762}} (\bibinfo {year} {2019})},\ \Eprint
  {https://arxiv.org/abs/1912.02762} {arXiv:1912.02762 [stat.ML]} \BibitemShut
  {NoStop}%
\bibitem [{Note4()}]{Note4}%
  \BibitemOpen
  \bibinfo {note} {\protect \href
  {https://github.com/jack-heinzel/population-error}{https://github.com/jack-heinzel/population-error}}\BibitemShut
  {NoStop}%
\bibitem [{\citenamefont {Talbot}\ \emph {et~al.}(2025)\citenamefont {Talbot},
  \citenamefont {Farah}, \citenamefont {Galaudage}, \citenamefont {Golomb},\
  and\ \citenamefont {Tong}}]{Talbot2025}%
  \BibitemOpen
  \bibfield  {author} {\bibinfo {author} {\bibfnamefont {C.}~\bibnamefont
  {Talbot}}, \bibinfo {author} {\bibfnamefont {A.}~\bibnamefont {Farah}},
  \bibinfo {author} {\bibfnamefont {S.}~\bibnamefont {Galaudage}}, \bibinfo
  {author} {\bibfnamefont {J.}~\bibnamefont {Golomb}},\ and\ \bibinfo {author}
  {\bibfnamefont {H.}~\bibnamefont {Tong}},\ }\href
  {https://doi.org/10.21105/joss.07753} {\bibfield  {journal} {\bibinfo
  {journal} {Journal of Open Source Software}\ }\textbf {\bibinfo {volume}
  {10}},\ \bibinfo {pages} {7753} (\bibinfo {year} {2025})},\ \Eprint
  {https://arxiv.org/abs/2409.14143} {arXiv:2409.14143 [astro-ph.IM]}
  \BibitemShut {NoStop}%
\bibitem [{\citenamefont {Varma}\ \emph {et~al.}(2019)\citenamefont {Varma},
  \citenamefont {Field}, \citenamefont {Scheel}, \citenamefont {Blackman},
  \citenamefont {Gerosa}, \citenamefont {Stein}, \citenamefont {Kidder},\ and\
  \citenamefont {Pfeiffer}}]{Varma:2019csw}%
  \BibitemOpen
  \bibfield  {author} {\bibinfo {author} {\bibfnamefont {V.}~\bibnamefont
  {Varma}}, \bibinfo {author} {\bibfnamefont {S.~E.}\ \bibnamefont {Field}},
  \bibinfo {author} {\bibfnamefont {M.~A.}\ \bibnamefont {Scheel}}, \bibinfo
  {author} {\bibfnamefont {J.}~\bibnamefont {Blackman}}, \bibinfo {author}
  {\bibfnamefont {D.}~\bibnamefont {Gerosa}}, \bibinfo {author} {\bibfnamefont
  {L.~C.}\ \bibnamefont {Stein}}, \bibinfo {author} {\bibfnamefont {L.~E.}\
  \bibnamefont {Kidder}},\ and\ \bibinfo {author} {\bibfnamefont {H.~P.}\
  \bibnamefont {Pfeiffer}},\ }\href
  {https://doi.org/10.1103/PhysRevResearch.1.033015} {\bibfield  {journal}
  {\bibinfo  {journal} {Phys. Rev. Research.}\ }\textbf {\bibinfo {volume}
  {1}},\ \bibinfo {pages} {033015} (\bibinfo {year} {2019})},\ \Eprint
  {https://arxiv.org/abs/1905.09300} {arXiv:1905.09300 [gr-qc]} \BibitemShut
  {NoStop}%
\bibitem [{\citenamefont {Pratten}\ \emph {et~al.}(2021)\citenamefont {Pratten}
  \emph {et~al.}}]{Pratten:2020ceb}%
  \BibitemOpen
  \bibfield  {author} {\bibinfo {author} {\bibfnamefont {G.}~\bibnamefont
  {Pratten}} \emph {et~al.},\ }\href
  {https://doi.org/10.1103/PhysRevD.103.104056} {\bibfield  {journal} {\bibinfo
   {journal} {Phys. Rev. D}\ }\textbf {\bibinfo {volume} {103}},\ \bibinfo
  {pages} {104056} (\bibinfo {year} {2021})},\ \Eprint
  {https://arxiv.org/abs/2004.06503} {arXiv:2004.06503 [gr-qc]} \BibitemShut
  {NoStop}%
\bibitem [{\citenamefont {Colleoni}\ \emph {et~al.}(2025)\citenamefont
  {Colleoni}, \citenamefont {Vidal}, \citenamefont {Garc{\'\i}a-Quir{\'o}s},
  \citenamefont {Ak{\c{c}}ay},\ and\ \citenamefont {Bera}}]{Colleoni:2024knd}%
  \BibitemOpen
  \bibfield  {author} {\bibinfo {author} {\bibfnamefont {M.}~\bibnamefont
  {Colleoni}}, \bibinfo {author} {\bibfnamefont {F.~A.~R.}\ \bibnamefont
  {Vidal}}, \bibinfo {author} {\bibfnamefont {C.}~\bibnamefont
  {Garc{\'\i}a-Quir{\'o}s}}, \bibinfo {author} {\bibfnamefont {S.}~\bibnamefont
  {Ak{\c{c}}ay}},\ and\ \bibinfo {author} {\bibfnamefont {S.}~\bibnamefont
  {Bera}},\ }\href {https://doi.org/10.1103/PhysRevD.111.104019} {\bibfield
  {journal} {\bibinfo  {journal} {Phys. Rev. D}\ }\textbf {\bibinfo {volume}
  {111}},\ \bibinfo {pages} {104019} (\bibinfo {year} {2025})},\ \Eprint
  {https://arxiv.org/abs/2412.16721} {arXiv:2412.16721 [gr-qc]} \BibitemShut
  {NoStop}%
\bibitem [{\citenamefont {Abac}\ \emph
  {et~al.}(2026{\natexlab{e}})\citenamefont {Abac} \emph
  {et~al.}}]{ligo_scientific_collaboration_2026_19500052}%
  \BibitemOpen
  \bibfield  {author} {\bibinfo {author} {\bibfnamefont {A.~G.}\ \bibnamefont
  {Abac}} \emph {et~al.} (\bibinfo {collaboration} {LIGO Scientific, Virgo,
  KAGRA}),\ }\href {https://doi.org/10.5281/zenodo.19500052}
  {10.5281/zenodo.19500052} (\bibinfo {year} {2026}{\natexlab{e}})\BibitemShut
  {NoStop}%
\bibitem [{\citenamefont {{Speagle}}(2020)}]{2020MNRAS.493.3132S}%
  \BibitemOpen
  \bibfield  {author} {\bibinfo {author} {\bibfnamefont {J.~S.}\ \bibnamefont
  {{Speagle}}},\ }\href {https://doi.org/10.1093/mnras/staa278} {\bibfield
  {journal} {\bibinfo  {journal} {Mon. Not. Roy. Astron. Soc.}\ }\textbf
  {\bibinfo {volume} {493}},\ \bibinfo {pages} {3132} (\bibinfo {year}
  {2020})},\ \Eprint {https://arxiv.org/abs/1904.02180} {arXiv:1904.02180
  [astro-ph.IM]} \BibitemShut {NoStop}%
\bibitem [{\citenamefont {Koposov}\ \emph {et~al.}(2024)\citenamefont {Koposov}
  \emph {et~al.}}]{sergey_koposov_2024_12537467}%
  \BibitemOpen
  \bibfield  {author} {\bibinfo {author} {\bibfnamefont {S.}~\bibnamefont
  {Koposov}} \emph {et~al.},\ }\href {https://doi.org/10.5281/zenodo.12537467}
  {\bibinfo {title} {joshspeagle/dynesty: v2.1.4}} (\bibinfo {year}
  {2024})\BibitemShut {NoStop}%
\bibitem [{\citenamefont {Ashton}\ \emph {et~al.}(2019)\citenamefont {Ashton}
  \emph {et~al.}}]{bilby_paper}%
  \BibitemOpen
  \bibfield  {author} {\bibinfo {author} {\bibfnamefont {G.}~\bibnamefont
  {Ashton}} \emph {et~al.},\ }\href {https://doi.org/10.3847/1538-4365/ab06fc}
  {\bibfield  {journal} {\bibinfo  {journal} {Astrophys. J. Suppl.}\ }\textbf
  {\bibinfo {volume} {241}},\ \bibinfo {pages} {27} (\bibinfo {year} {2019})},\
  \Eprint {https://arxiv.org/abs/1811.02042} {arXiv:1811.02042 [astro-ph.IM]}
  \BibitemShut {NoStop}%
\bibitem [{\citenamefont {Talbot}\ \emph {et~al.}(2026)\citenamefont {Talbot}
  \emph {et~al.}}]{colm_talbot_2026_18788906}%
  \BibitemOpen
  \bibfield  {author} {\bibinfo {author} {\bibfnamefont {C.}~\bibnamefont
  {Talbot}} \emph {et~al.},\ }\href {https://doi.org/10.5281/zenodo.18788906}
  {\bibinfo {title} {bilby-dev/bilby: v2.8.0}} (\bibinfo {year}
  {2026})\BibitemShut {NoStop}%
\bibitem [{\citenamefont {Kimball}\ \emph {et~al.}(2021)\citenamefont {Kimball}
  \emph {et~al.}}]{Kimball:2020qyd}%
  \BibitemOpen
  \bibfield  {author} {\bibinfo {author} {\bibfnamefont {C.}~\bibnamefont
  {Kimball}} \emph {et~al.},\ }\href {https://doi.org/10.3847/2041-8213/ac0aef}
  {\bibfield  {journal} {\bibinfo  {journal} {Astrophys. J. Lett.}\ }\textbf
  {\bibinfo {volume} {915}},\ \bibinfo {pages} {L35} (\bibinfo {year}
  {2021})},\ \Eprint {https://arxiv.org/abs/2011.05332} {arXiv:2011.05332
  [astro-ph.HE]} \BibitemShut {NoStop}%
\bibitem [{\citenamefont {Galaudage}\ \emph {et~al.}(2021)\citenamefont
  {Galaudage} \emph {et~al.}}]{Galaudage:2021rkt}%
  \BibitemOpen
  \bibfield  {author} {\bibinfo {author} {\bibfnamefont {S.}~\bibnamefont
  {Galaudage}} \emph {et~al.},\ }\href
  {https://doi.org/10.3847/2041-8213/ac2f3c} {\bibfield  {journal} {\bibinfo
  {journal} {Astrophys. J. Lett.}\ }\textbf {\bibinfo {volume} {921}},\
  \bibinfo {pages} {L15} (\bibinfo {year} {2021})},\ \bibinfo {note} {[Erratum:
  Astrophys.J.Lett. 936, L18 (2022), Erratum: Astrophys.J. 936, L18 (2022)]},\
  \Eprint {https://arxiv.org/abs/2109.02424} {arXiv:2109.02424 [gr-qc]}
  \BibitemShut {NoStop}%
\bibitem [{\citenamefont {Callister}\ \emph {et~al.}(2022)\citenamefont
  {Callister}, \citenamefont {Miller}, \citenamefont {Chatziioannou},\ and\
  \citenamefont {Farr}}]{Callister:2022qwb}%
  \BibitemOpen
  \bibfield  {author} {\bibinfo {author} {\bibfnamefont {T.~A.}\ \bibnamefont
  {Callister}}, \bibinfo {author} {\bibfnamefont {S.~J.}\ \bibnamefont
  {Miller}}, \bibinfo {author} {\bibfnamefont {K.}~\bibnamefont
  {Chatziioannou}},\ and\ \bibinfo {author} {\bibfnamefont {W.~M.}\
  \bibnamefont {Farr}},\ }\href {https://doi.org/10.3847/2041-8213/ac847e}
  {\bibfield  {journal} {\bibinfo  {journal} {Astrophys. J. Lett.}\ }\textbf
  {\bibinfo {volume} {937}},\ \bibinfo {pages} {L13} (\bibinfo {year}
  {2022})},\ \Eprint {https://arxiv.org/abs/2205.08574} {arXiv:2205.08574
  [astro-ph.HE]} \BibitemShut {NoStop}%
\bibitem [{\citenamefont {Tong}\ \emph {et~al.}(2022)\citenamefont {Tong},
  \citenamefont {Galaudage},\ and\ \citenamefont {Thrane}}]{Tong:2022iws}%
  \BibitemOpen
  \bibfield  {author} {\bibinfo {author} {\bibfnamefont {H.}~\bibnamefont
  {Tong}}, \bibinfo {author} {\bibfnamefont {S.}~\bibnamefont {Galaudage}},\
  and\ \bibinfo {author} {\bibfnamefont {E.}~\bibnamefont {Thrane}},\ }\href
  {https://doi.org/10.1103/PhysRevD.106.103019} {\bibfield  {journal} {\bibinfo
   {journal} {Phys. Rev. D}\ }\textbf {\bibinfo {volume} {106}},\ \bibinfo
  {pages} {103019} (\bibinfo {year} {2022})},\ \Eprint
  {https://arxiv.org/abs/2209.02206} {arXiv:2209.02206 [astro-ph.HE]}
  \BibitemShut {NoStop}%
\end{thebibliography}%

\clearpage

\section{End Matter}

\textit{Pushforward Approximation}---We approximate the pushforward of a component spin (and mass ratio) population model, Eq.~\eqref{eq:pf},
with a parametric density $\pf$,
\begin{equation}
    \pf(\chib \mid \pfp) = G(\chie \mid \pfp) B(\chip \mid \pfp) \, ,
\end{equation}
where $\gamma$ are the parameters of the Beta distribution $B$ and generalized Gaussian $G$ \cite{subbotin1923law},
\begin{equation} \label{eq:gen-gauss}
    G(x) = \frac{\beta}{2 \alpha \Gamma(1 / \beta)} e^{-(|x - \mu|/\alpha)^{\beta}} \, .
\end{equation}
Here, $\Gamma$ is the Gamma function,
$\alpha, \beta > 0$ are shape parameters, and $\mu$ is the mean.
When $\beta = 2$ we recover the standard Gaussian distribution,
when $\beta = 1$ the Laplace distribution,
and in the limit that $\beta \rightarrow \infty$ the tophat distribution.
We also neglect conditioning the pushforward on $q$ which we found to have a negligible impact for the population models we considered.

To wit, at fixed $\lambda$
we minimize a \ac{MC} estimate of the \ac{KL} divergence \cite{kl}
between $\pf(\chib \mid \pfp)$ and the pushforward $p(\chib \mid \lambda)$ marginalized over $q$,
\begin{equation}
\begin{aligned}
    \mathrm{KL}(p , \pf) &= \int \dd \chib \,  p(\chib \mid \lambda) \ln \left( \frac{ p(\chib \mid \lambda) }{ \pf(\chib \mid \pfp) } \right) \\
    &\propto -\int \dd \chib \, p(\chib \mid \lambda) \ln \pf(\chib \mid \pfp) \\
    &\approx -\frac{1}{n} \sum_{i = 1}^n \ln \pf(\chib(s_i, q_i) \mid \pfp) \, .
\end{aligned}
\end{equation}
We draw $n = 10^4$ samples $s_i, q_i$ from the component spin population model $p(s, q \mid \lambda)$. 
We minimize $\mathrm{KL}(p , \pf)$ as a function of $\gamma$
using the \texttt{BFGS} solver implemented in \texttt{optimistix} \cite{optimistix2024}.

\begin{figure}[t]
    \centering
    \includegraphics[width=0.87\linewidth]{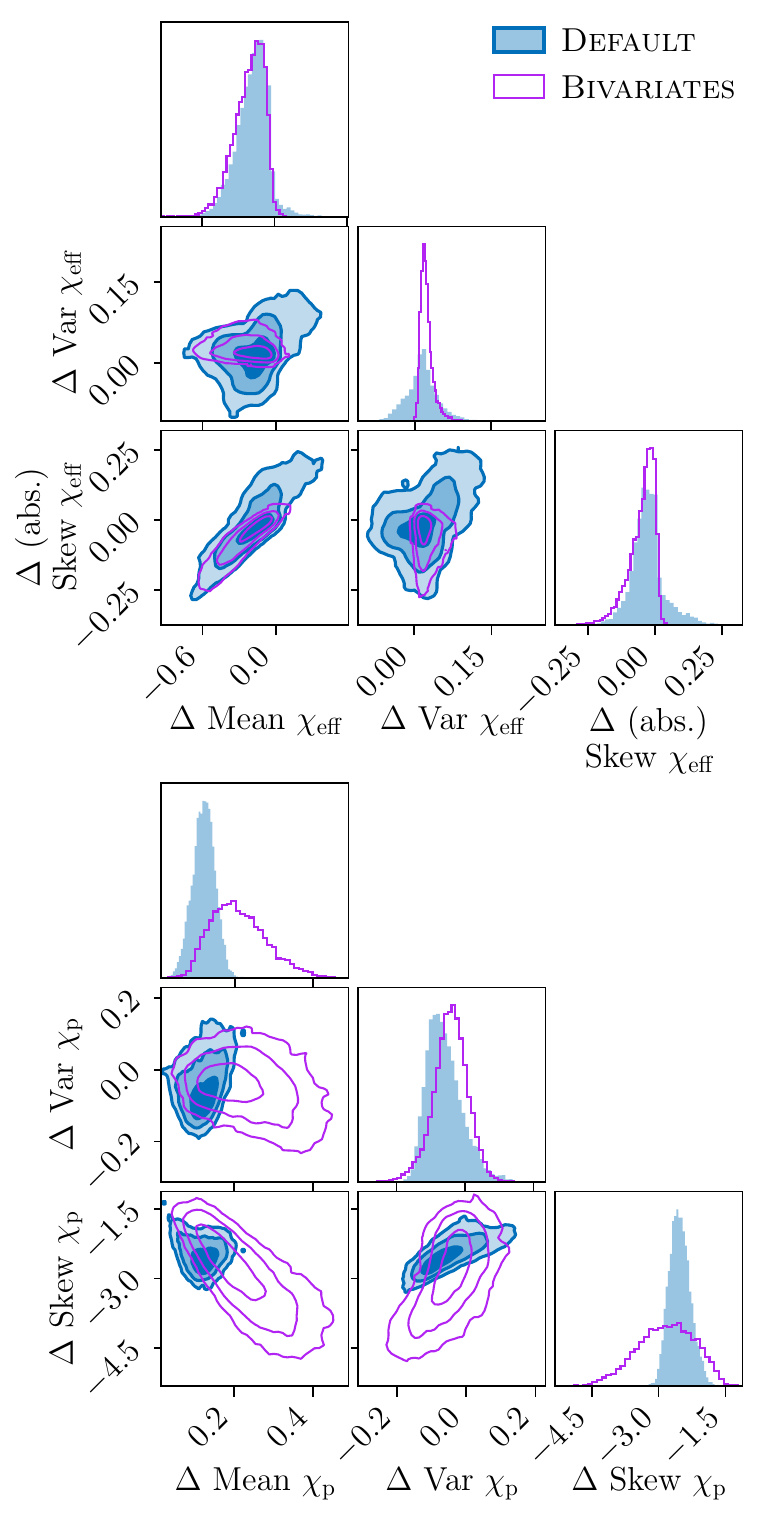}
    \caption{
    Residuals on the mean, variance, and skew
    of the effective spin pushforward approximation
    over the \moddef (blue) and \modbiv (purple) posteriors.
    Moments of \chie (\chip) are on top (bottom).
    Residuals are reported in units of the posterior standard deviation for each moment, except for the \chie skew which is an absolute difference.
    Contours enclose 50\%, 90\%, and 99\% credible regions.
    }
    \label{fig:skew-kurt}
\end{figure}

Our choice of $g$ does not capture the detailed shape
of the effective spin pushforward
under either the \moddef or \modbiv component spin models.
For example, the mixture of isotropic and Gaussian tilt distributions
in the \moddef model will induce skewness in the \chie pushforward \cite{Banagiri:2025dxo}.
Motivated by this example,
in Fig.~\ref{fig:skew-kurt}
we plot residuals on the mean, variance, and skew
of $p(\chib \mid \lambda)$ as well as $\pf(\chib \mid \pfp)$
over the \moddef and \modbiv posteriors.
Note that we estimate moments of the pushforward with \ac{MC} samples.
The mean, variance, and skew of the pushforward approximant
are calculated analytically from $\gamma$.
For each posterior sample,
we report the residual of the moment $\nu$
as $(\nu_{\pf} - \nu) / \sigma_\nu $
where $\nu_{\pf}$ is that same moment
under the pushforward approximant
and $\sigma_\nu$ is the posterior standard deviation on $\nu$.
Note that, by construction, the approximant \chie (generalized Gaussian) skew is always zero,
so we report the \chie skew residual as an absolute difference.
We find that the \chib means and variances are well approximated by $g$,
with relative errors of $\lesssim 0.5$ standard deviations.
However, there is non-zero \chie skew
which is not captured by the pushforward approximant.
Further, the \chip skew is underestimated by as much as $\sim3$ ($\sim 4.5$) standard deviations
for the \moddef (\modbiv) model.
Misestimation of the effective spin skews
may contribute to discrepancies
between
restricted and full posteriors
under both the \moddef and \modbiv models.

We did not optimize our choice of $g$
and have not tested how sensitive
our results are to this choice.
We initially estimated the pushforward density using Gaussian \acp{KDE}.
However, \acp{KDE} have the undesirable property that the $\chip$ density does not go to zero as $\chip \rightarrow 0$,
which motivated our choice of the Beta distribution for $\chip$.
Alternatively,
the pushforward could be approximated
with flexible density emulators conditioned on $\lambda$ and trained prior to inference,
like normalizing flows \cite{2019arXiv191202762P}.

\textit{Likelihood estimation}---We estimate the population log-likelihood via \ac{MC} integration.
During inference, we threshold the variance $\mathcal{V}(\lambda)$ of the log-likelihood estimator $\hat{p}(\{ d\} \mid \lambda)$ to be at most 1 \cite{Tiwari:2017ndi, Essick:2022ojx, Talbot:2023pex, Heinzel:2025ogf}.
The variance of the log-likelihood estimator changes
when the population model is changed.
When performing inference with the pushforward population model, we make two modifications to the variance threshold.
First, we impose a threshold of 1 on the variance of the likelihood under the \textit{component spin model},
not its pushforward,
to ensure a like-for-like comparison
between the
full and restricted posterior.
Second, we impose an additional threshold of 4 on the variance of the likelihood under the pushforward model for numerical stability.
In practice, the likelihood we use for analysis with the pushforward is,
\begin{equation}
    \hat{p}(\{ d \} \mid \lambda, \mathrm{eff}) \, \Theta( \mathcal{V}(\lambda \mid \mathrm{eff}) \leq 4 ) \, \Theta( \mathcal{V}(\lambda) \leq 1  ) \, ,
\end{equation}
where $\mathcal{V}(\lambda \mid \mathrm{eff})$ is the variance of $\hat{p}(\{ d \} \mid \lambda, \mathrm{eff})$
and $\Theta$ is the Heaviside step function.

\textit{Evidence uncertainty propagation}---Uncertainty in the log-likelihood estimator is $\lesssim 1$ over the population posteriors in both component spin analyses.
Here, we propagate this uncertainty into the Bayesian evidences and Bayes factors.
Following Ref.~\cite{Heinzel:2025ogf},
we define the unnormalized posterior estimator $\mathcal{L}(\lambda) \pi(\lambda) (1 + \delta)$ where $\mathcal{L}$ is the likelihood, $\pi$ is the prior, and $\delta(\lambda) = b(\lambda) + \varepsilon(\lambda)$.
The posterior $p(\lambda)$ has normalization (i.e. evidence) $Z = \int \dd \lambda \, \mathcal{L}(\lambda) \pi(\lambda)$.
Under \ac{MC} realizations,
the bias $b$ is fixed while $\varepsilon$  characterizes the zero-mean scatter of the log-likelihood estimator.
The evidence estimator is
\begin{equation}
    \hat{Z} = \int \dd \lambda \, \mathcal{L}(\lambda) \pi(\lambda) (1 + \delta(\lambda)) = Z \left[1  + \int \dd \lambda \, p(\lambda) \, \delta(\lambda) \right] \, ,
\end{equation}
with mean
\begin{align}
    \mean{\hat{Z}} = Z \left[ 1 + \int \dd \lambda \, p(\lambda) \mean{\delta(\lambda)} \right] \, ,
\end{align}
where $\mean{\cdot}$ denotes the mean under \ac{MC} realizations.
Similarly,
\begin{equation}
\begin{aligned}
    \mean{\hat{Z}^2} = Z^2 \bigg[ 1 + &2  \int \dd \lambda \, p(\lambda) \mean{\delta(\lambda)} \\
    &+ \int \dd \lambda \dd \lambda' \, p(\lambda)p(\lambda') \mean{\delta(\lambda)\delta(\lambda')} \bigg] \, .
\end{aligned}
\end{equation}
Then, the variance of the evidence estimator is
\begin{equation}
\begin{aligned}
    \Var{\hat{Z}} &= \mean{\hat{Z}^2} - \mean{\hat{Z}}^2 \\
    &= Z^2\int \dd \lambda \dd \lambda' \, p(\lambda) p(\lambda') \left[ \mean{\delta(\lambda)\delta(\lambda')} - \mean{\delta(\lambda)} \mean{\delta(\lambda')} \right] \\
    &= Z^2\int \dd \lambda \dd \lambda' \, p(\lambda) p(\lambda') \mean{\varepsilon(\lambda)\varepsilon(\lambda')} \, .
\end{aligned}
\end{equation}
The variance of the log evidence estimator is
\begin{equation} \label{eq:varlnz-approx}
\begin{aligned}
    &\Var{\ln \hat{Z}} \approx \mean{\hat{Z}}^{-2} \Var{\hat{Z}} = \frac{ Z^2 \int \dd \lambda \dd \lambda' \, p(\lambda) p(\lambda') \mean{\varepsilon(\lambda)\varepsilon(\lambda')} }{ Z^2 \left[ 1 + \int \dd \lambda \, p(\lambda) \, \mean{\delta(\lambda)} \right]^2 } \\
    &\approx \left[ 1 - 2 \int \dd \lambda \, p(\lambda) \mean{\delta(\lambda)} \right] \int \dd \lambda \dd \lambda' \, p(\lambda) p(\lambda') \mean{\varepsilon(\lambda)\varepsilon(\lambda')} \\
    &\approx Z^{-2} \Var{\hat{Z}} \, ,
\end{aligned}
\end{equation}
where in the second line we expand the denominator to $\mathcal{O}(\delta)$
and in the third line we truncate terms of $\mathcal{O}(\delta^3)$ following Ref.~\cite{Heinzel:2025ogf}.
We notate the \ac{MC} covariance of the log-likelihood estimator as $\mean{\varepsilon(\lambda) \varepsilon(\lambda')} \equiv \hat{C}_{\ln \mathcal{L}}(\lambda, \lambda')$ \red{with $\hat{C}_{\ln \mathcal{L}}(\lambda, \lambda) = \mathcal{V}(\lambda)$}.
We can estimate Eq.~\eqref{eq:varlnz-approx} as
\begin{equation} \label{eq:varlnz-approx-coarse}
    \Var{\ln \hat{Z}} \approx \frac{1}{N^2} \sum_i \sum_j \hat{C}_{\ln \mathcal{L}}(\lambda_i, \lambda_j) \, .
\end{equation}
Here, note that $\lambda_{i,j}$ are drawn from the posterior estimator.

Just as log-likelihood estimators at different $\lambda$
may be covariate because they are estimated with the same \ac{MC} samples,
evidence estimators from two different models
may be covariate.
We use subscripts ``1'' and ``2'' to differentiate the posterior, log-likelihood scatter, and evidence
evaluated under different models.
The evidence covariance is
\begin{equation}
    \mathrm{Cov}(\hat{Z}_1, \hat{Z_2}) = Z_1 Z_2 \int \dd \lambda \, \dd \lambda' \, p_1(\lambda) p_2(\lambda') \mean{ \varepsilon_1(\lambda)\varepsilon_2(\lambda') } \, , \\
\end{equation}
where an unprimed (primed) $\lambda$ denotes parameters of model 1 (2).
The covariance of log evidence estimators is
\begin{equation} \label{eq:varlncov-approx}
\begin{aligned}
    \mathrm{Cov}(\ln \hat{Z}_1, \ln \hat{Z}_2) &\approx \mean{\hat{Z}_1}^{-1} \mean{\hat{Z}_2}^{-1} \mathrm{Cov}(\hat{Z}_1, \hat{Z_2)} \\
    &\approx Z_1^{-1} Z_2^{-1} \mathrm{Cov}(\hat{Z}_1, \hat{Z}_2) \\
    &\approx \frac{1}{N_1} \frac{1}{N_2} \sum_{i = 1}^{N_1} \sum_{j = 1}^{N_2} \hat{C}_{\ln \mathcal{L}}(\lambda_i, \lambda_j') \, ,
\end{aligned}
\end{equation}
where from the first to second lines
we expand $\mean{\hat{Z}_{1,2}}^{-1}$ to $\mathcal{O}(\delta)$ and then truncate the entire expression at $\mathcal{O}(\delta^3)$
as in Eq.~\eqref{eq:varlnz-approx}.

The variance of the log Bayes factor between model 1 and 2 is $\Var{\ln {\Delta \hat{Z}}} = \Var{\ln \hat{Z}_1} + \Var{\ln \hat{Z}_2} - 2 \, \mathrm{Cov}(\ln \hat{Z}_1, \ln \hat{Z}_2) $.
We compute Eqs.~\eqref{eq:varlnz-approx-coarse} and \eqref{eq:varlncov-approx} with functions in the \texttt{population-error} package \footnote{ \href{https://github.com/jack-heinzel/population-error}{https://github.com/jack-heinzel/population-error}}.
We find that \ac{MC} uncertainty in \lnb
is larger than uncertainty from nested sampling.

\clearpage

\section{Supplemental Material}

\begin{table}[h]
    \centering
    \begin{tabular}{ l c c }
        \toprule
        \toprule
        \textbf{Parameter} & \textbf{Symbol} & \textbf{Prior} \\
        \midrule
        Primary magnitude location & $\mu_{a_1}$ & $[0, 1]$ \\
        Secondary magnitude location & $\mu_{a_2}$ & $[0, 1]$ \\
        Primary magnitude scale & $\sigma_{a_1}$ & $[0.005, 1]$ \\
        Secondary magnitude scale & $\sigma_{a_2}$ & $[0.005, 1]$ \\
        Magnitude correlation parameter & $\rho_{a}$ & $[-0.99, 0.99]$ \\
        \midrule
        Primary tilt location & $\mu_{t_1}$ & $[-1, 1]$ \\
        Secondary tilt location & $\mu_{t_2}$ & $[-1, 1]$ \\
        Primary tilt scale & $\sigma_{t_1}$ & $[0.1, 4]$ \\
        Secondary tilt scale & $\sigma_{t_2}$ & $[0.1, 4]$ \\
        Tilt correlation parameter & $\rho_{t}$ & $[-0.99, 0.99]$ \\
        \bottomrule
        \bottomrule
    \end{tabular}
    \caption{
    Priors on the parameters specifying the \modbiv distribution of
    spin magnitudes (top) and tilts (bottom).
    We denote a uniform prior between $l$ and $u$ as $[l, u]$.
    }
    \label{tab:biv-priors}
\end{table}

\textit{Population models and priors}---Both the \moddef and \modbiv models
adopt the preferred parametric models
for mass and redshift from Ref.~\cite{LIGOScientific:2026ctl}.
We also adopt the same priors on the mass and redshift model parameters as that work.
Additionally, 
in both the \moddef and \modbiv models
the spin magnitudes and tilts are independently distributed,
$p(s \mid \lambda) = p(a_{1,2}) p(\ct_{1,2})$.
Here, we define $p(a_{1,2})$ and $p(\ct_{1,2})$ for each model.

The \moddef model assumes primary and secondary spin magnitudes are \ac{IID} according to truncated Gaussians,
\begin{equation}
    p(a_{1,2}) = \mathcal{N}_{[0, 1]}(a_1 \mid \mu_a, \sigma_a) \, \mathcal{N}_{[0, 1]}(a_2 \mid \mu_a, \sigma_a) \,  ,
\end{equation}
where $\mathcal{N}_{[l, u]}(\cdot)$ denotes a truncated Gaussian on $[l, u]$
and $\mu_a$ and $\sigma_a$ are the location and scale, respectively.
The spin tilts are modeled as a mixture of an isotropic distribution and a truncated Gaussian,
\begin{equation}
\begin{aligned}
    &p(\ct_{1,2}) = \bigg[ \frac{1 - \xi}{4} +  \\
    &+ \xi \, \mathcal{N}_{[-1, 1]}(\ct_1 \mid \mu_t, \sigma_t) \mathcal{N}_{[-1, 1]}(\ct_2 \mid \mu_t, \sigma_t) \bigg] \, ,
\end{aligned}
\end{equation}
with branching ratio $\xi$.
The location $\mu_t$ and scale $\sigma_t$
of the truncated Gaussians
are the same for primary and secondary tilts.
Priors on the \moddef spin model parameters
follow Tab.~6 of Ref.~\cite{LIGOScientific:2026ctl}.

The \modbiv model allows
correlations between primary and secondary spins,
although magnitudes and tilts are still treated independently.
To wit,
\begin{equation}
    p(a_{1,2}) = \mathcal{N}_{[0,1] \times [0,1]}(a_{1,2} \mid  \bm{\mu}_a, \bm{\Sigma}_a) \, ,
\end{equation}
where $\mathcal{N}_{[0,1] \times [0,1]}$ is a bivariate Gaussian
truncated to the box $[0,1] \times [0,1]$
with location $\bm{\mu}_a = \{ \mu_{a_1}, \mu_{a_2} \}$.
We use the bivariate Gaussian implemented in \texttt{gwpopulation} \cite{Talbot2025},
where the covariance matrix $\bm{\Sigma}_a$ has the form
\begin{equation}
    \bm{\Sigma}_a = \frac{1}{1 + \rho_a} \begin{bmatrix}
    \sigma_{a_1}^2 & \rho_a \, \sigma_{a_1} \sigma_{a_2} \\
    \rho_a \, \sigma_{a_1} \sigma_{a_2} & \sigma_{a_2}^2
    \end{bmatrix} \, ,
\end{equation}
where $\sigma_{a_{1,2}}$ are the marginal scales
and $\rho \in (-1, 1)$
is the correlation parameter.
Similarly,
\begin{equation}
    p(\ct_{1,2}) = \mathcal{N}_{[-1,1] \times [-1,1]}(\ct_{1,2} \mid  \bm{\mu}_t, \bm{\Sigma}_t) \, ,
\end{equation}
where the locations and covariance matrix take the same form as $\bm{\mu_a}, \bm{\Sigma}_a$ while swapping the labels $a \leftrightarrow t$.
Priors on the \modbiv spin parameters are given in Tab.~\ref{tab:biv-priors}.

\begin{figure}
    \centering
    \includegraphics[width=0.99\linewidth]{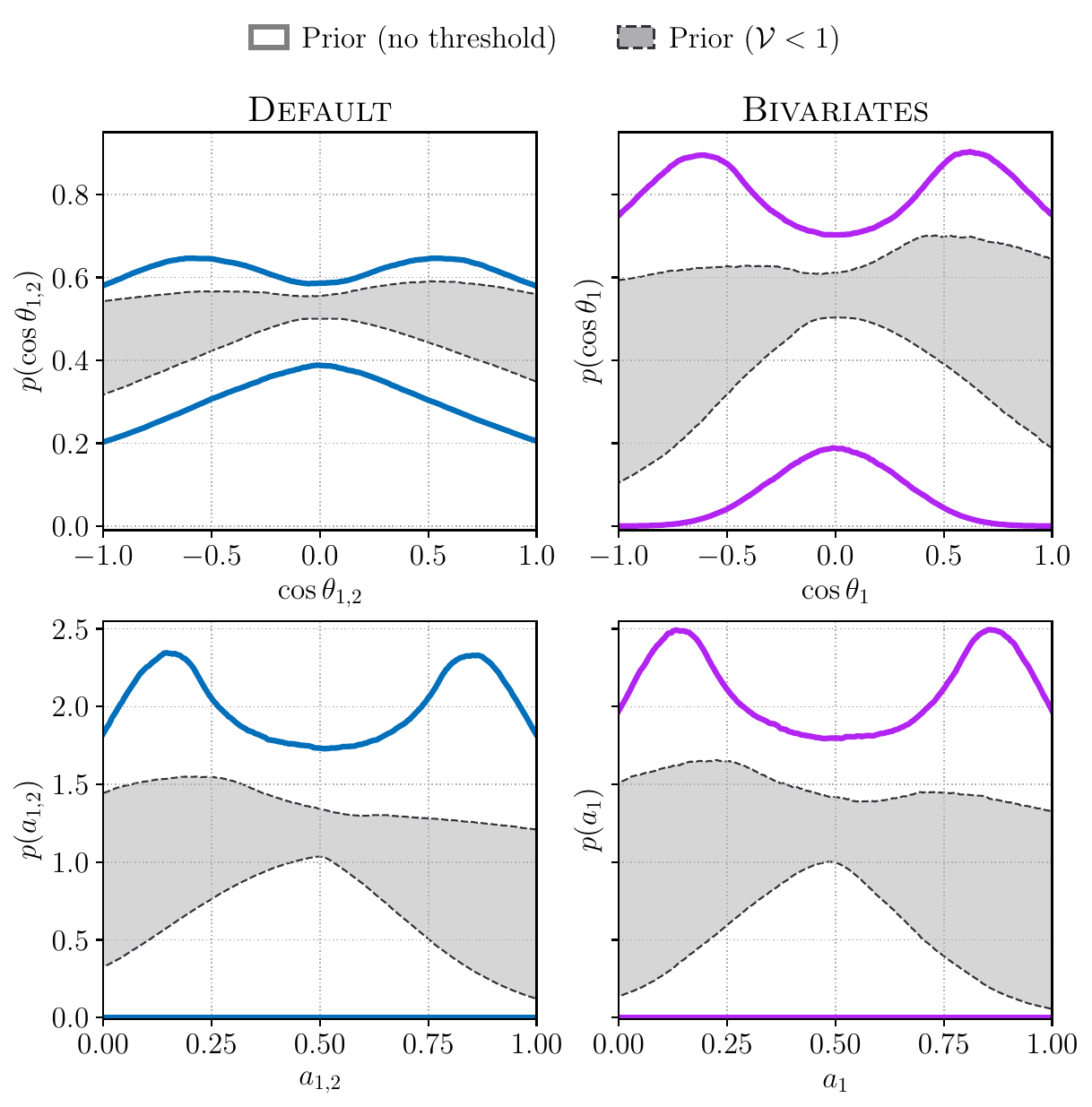}
    \caption{Population priors for spin tilts (top) and magnitudes (bottom) for the \moddef and \modbiv models. Lines denote the 90\% credible region of the prior;
    shading denotes the same region
    after imposing a maximum log-likelihood variance $\mathcal{V}$ of 1.
    Marginal priors on secondary spins under the \modbiv model are identical to the primary spin priors shown in the right column.
    }
    \label{fig:priors-variance-comparison}
\end{figure}

\red{
We plot population priors for the \moddef (\modbiv) model in the left (right) column of Fig.~\ref{fig:priors-variance-comparison}.
We also show priors 
after excluding samples
for which the log-likelihood \ac{MC} variance is above 1 (cf. the End Matter).
This threshold asymmetrically modifies the spin population prior,
preferring
spin tilts aligned with the binary orbit
or small spin magnitudes.
While slight,
this preference qualitatively reflects
the simulated population
used to estimate selection effects
(cf. Fig.~5 of Ref.~\cite{Essick:2025zed}).
}

\textit{Data and analysis settings}---We analyze \acp{BBH} in the (cumulative) fifth Gravitational-Wave Transient Catalog (GWTC-5).
We include events from the \ac{LVK}'s third observing run (O3) and onward with false-alarm rates $< 1\,\mathrm{yr}^{-1}$;
we include events from the first (O1) and second (O2) observing runs
with signal-to-noise ratios $> 10$.
For all events we use parameter estimation obtained
with the \textsc{NRSur7dq4} waveform approximant \cite{Varma:2019csw}
when available
and \textsc{IMRPhenomXPHM} otherwise \cite{Pratten:2020ceb, Colleoni:2024knd},
resulting in 6,193 posterior samples per event.
Selection effects are estimated with public \ac{LVK} data products \cite{Essick:2025zed, ligo_scientific_collaboration_2026_19500052}.

All analyses are performed with \texttt{dynesty} \cite{2020MNRAS.493.3132S, sergey_koposov_2024_12537467} wrapped by \texttt{bilby} \cite{bilby_paper, colm_talbot_2026_18788906}.
We use 1000 live points, and the ``acceptance walk'' with \texttt{naccept = 5} to estimate likelihood isocontours.
We use the population likelihood implemented in \texttt{gwpopulation}.

\textit{\red{Toy example}}---Assume that spin tilts are distributed independently of spin magnitudes and mass ratio;
then, from Eq.~\eqref{eq:chieff-def} and by linearity of expectations,
\begin{equation}
    \mean{\chie} = \mean{ \frac{a_1}{1 + q} } \mean{\ct_1} + \mean{ \frac{q\, a_2}{1 + q} } \mean{\ct_2} \, ,
\end{equation}
and
\begin{equation}
\begin{aligned}
    \mean{\chie^2} = & \bigg[ \mean{\frac{a_1^2}{(1 + q)^2}} \mean{\ct_1^2} \\
    &+ \mean{ \frac{q^2 a_2^2}{ (1 + q)^2 } } \mean{\ct_2^2} \\
    &+ 2 \mean{\frac{q \, a_1a_2}{(1 + q)^2}} \mean{\ct_1 \ct_2} \bigg] \, ,
\end{aligned}
\end{equation}
where $\mean{\cdot}$ denotes expectations over the astrophysical population.
Define mass-weighted spin magnitudes $x = a_1 / (1 + q)$ and $y = q \, a_2 / (1 + q)$.
Further, assume that primary and secondary tilts are \ac{NID}.
Then,
\begin{equation}
    \mean{\chie} = \mean{\ct}\mean{x + y} \, ,\label{eq:mean-chie}
\end{equation}
and,
\begin{equation} \label{eq:var-chie}
\begin{aligned}
    \Var{\chie} = \bigg[ &\mean{\ct^2} \mean{x^2 + y^2} + 2 \mean{xy} \mean{\ct_1 \ct_2} \\ &- \mean{\ct}^2 \mean{x + y}^2 \bigg] \, .
\end{aligned}
\end{equation}
Note that the Pearson correlation between $\ct_1$ and $\ct_2$ is at minimum $-1$, so
\begin{equation}
    \mean{\ct_1 \ct_2} \geq \mean{\ct}^2 - \Var{\ct} \, .
\end{equation}
We apply this inequality to Eq.~\eqref{eq:var-chie};
we also replace $\mean{\ct^2}$ in Eq.~\eqref{eq:var-chie}
with $\Var{\ct} + \mean{\ct}^2$, yielding
\begin{equation}
\begin{aligned}
    \Var{&\chie} \geq \bigg(\Var{\ct} \left[ \mean{x^2 + y^2} - 2 \mean{xy} \right] \\
    &+ \mean{\ct}^2 \left[ \mean{x^2 + y^2} + 2 \mean{xy} - \mean{x + y}^2 \right] \bigg) \, .
\end{aligned}
\end{equation}
Note that $\mean{(x-y)^2} = \mean{x^2 + y^2} - 2\mean{xy}$.
Also,
\begin{equation}
\begin{aligned}
    \Var{(x + y)} &= \bigg[ \mean{x^2} - \mean{x}^2 + \mean{y^2} - \mean{y}^2 \\
    &\quad \quad \quad+ 2\mean{xy} - 2\mean{x}\mean{y} \bigg] \\
    &= \mean{x^2 + y^2} + 2 \mean{xy} - \mean{x + y}^2 \, ,
\end{aligned}
\end{equation}
which allows us to write
\begin{equation} \label{eq:clean-var-chie-bound}
    \Var{\chie} \geq \Var{\ct} \mean{(x - y)^2} + \mean{\ct}^2 \Var{(x + y)} \, .
\end{equation}
Importantly, all terms on the right-hand side of Eq.~\eqref{eq:clean-var-chie-bound} are non-negative.

Consider a delta function \chie population  located at $\mean{\chie} = 0$. 
Then, $\Var{\chie} = 0$.
Considering both Eqs.~\eqref{eq:mean-chie} and \eqref{eq:clean-var-chie-bound}, we require
\begin{align}
    \mean{\ct} \mean{x + y} &= 0 \label{eq:req1} \, , \\
    \mean{\ct}^2 \Var{(x + y)} &= 0 \, , \label{eq:req2} \\
    \Var{\ct} \mean{(x - y)^2} &= 0 \, . \label{eq:req3} 
\end{align}
However, note that Eqs.~\eqref{eq:req1} and \eqref{eq:req2}
are satisfied under the same two cases:
$\mean{\ct} = 0$ or $\mean{x + y} = 0$.
In particular, Eq.~\eqref{eq:req2} is satisfied with $\Var{(x + y)} = 0$ in the latter case since $\mean{x + y} = 0$ implies all $x = y = 0$ given that $x, y \geq 0$;
any draw $x, y > 0$ would drag $\mean{x + y}$ above zero.

Further, Eq.~\eqref{eq:req3} is satisfied
when $\Var{\ct} = 0$ or $\mean{(x - y)^2} = 0$.
The latter case implies $x = y$ since $x, y \geq 0$.
Therefore, the cases that satisfy Eqs.~\eqref{eq:req1}, \eqref{eq:req2}, and \eqref{eq:req3} reduce to
\begin{equation}
\begin{aligned}
    &\Var{\ct}= 0 \text{ and } \mean{\ct} = 0 \text{, or, } \\
    &x = y \text{ and either } \mean{\ct} = 0 \text{ or } x = y = 0 \, .
\end{aligned}
\end{equation}
The first case means all population draws have $\ct_1 = \ct_2 = 0$.
In the second case, if we further specify that spin magnitudes are independent of mass ratio,
we have that all $a_1 = a_2 = 0$ since $q > 0$.
Note that both cases are formally excluded by the population priors employed in this work 
as neither the $\ct_{1,2}$ nor $a_{1,2}$ distributions can have zero variance.
Zero width populations are also excluded by population log-likelihood \ac{MC} variance thresholds.
These caveats apply to many \ac{GW} population inferences,
although see, e.g., Refs.~\cite{Kimball:2020qyd, Galaudage:2021rkt, Callister:2022qwb, Mould:2022xeu, Tong:2022iws, Hussain:2024qzl}
which model narrow component spin distributions.

\end{document}